\documentclass[twocolumn, times]{aastex631}
\usepackage{booktabs}
\usepackage{marginnote}

\usepackage{soul}

\makeatletter
\DeclareRobustCommand{\HI}{%
  \mbox{H\,\check@mathfonts\fontsize\sf@size\z@\selectfont I}%
}
\DeclareRobustCommand{\relation}{%
  \mbox{$M_{\rm HI,grp}$--$M_{\rm halo}$ relation}%
}
\DeclareRobustCommand{\fsmgrgrp}{%
  \mbox{$\rm FSMGR_{\rm grp}$}%
}

\DeclareRobustCommand{\rosat}{%
  \mbox{\textsl{ROSAT}}%
}

\begin{document}

\title{{The RESOLVE and ECO G3 Initiative: }Drivers of \HI{} Content and X-ray Emission in Galaxy Groups}

\correspondingauthor{Zackary L. Hutchens}
\email{zhutchen@live.unc.edu}
\author[0000-0002-8574-5495]{Zackary L. Hutchens}
\affiliation{Department of Physics \& Astronomy, University of North Carolina at Chapel Hill, 120 E. Cameron Ave., Chapel Hill, NC 27599, USA}
\affiliation{{Department of Physics \& Astronomy, University of North Carolina Asheville, 1 University Heights, Asheville, NC 28804, USA}}

\author[0000-0002-3378-6551]{Sheila J. Kannappan}
\affiliation{Department of Physics \& Astronomy, University of North Carolina at Chapel Hill, 120 E. Cameron Ave., Chapel Hill, NC 27599, USA}

\author[0000-0001-9662-9089]{Kelley M. Hess}
\affiliation{Department of Space, Earth and Environment, Chalmers University of Technology, Onsala Space Observatory, 43992 Onsala, Sweden}
\affiliation{ASTRON, the Netherlands Institute for Radio Astronomy, Postbus 2, 7990 AA, Dwingeloo, The Netherlands}
\affiliation{Instituto de Astrof\'{i}sica de Andaluc\'{i}a (CSIC), Glorieta de la Astronom\'{i}a, 18008 Granada, Spain}

\author[0000-0002-7892-396X]{Andrew J. Baker}
\affiliation{Department of Physics and Astronomy, Rutgers, the State University of New Jersey, 136 Frelinghuysen Road, Piscataway, NJ 08854-8019, USA}
\affiliation{Department of Physics and Astronomy, University of the Western Cape, Robert Sobukwe Road, Bellville 7535, South Africa}

\author[0000-0001-5880-0703]{Ming Sun}
\affiliation{Department of Physics and Astronomy, The University of Alabama in Huntsville, 301 Sparkman Drive, Huntsville, AL 35899, USA}

\author{Derrick S. Carr}
\author{Kathleen D. Eckert}
\affiliation{Department of Physics \& Astronomy, University of North Carolina at Chapel Hill, 120 E. Cameron Ave., Chapel Hill, NC 27599, USA}

\author{David V. Stark}
\affiliation{Space Telescope Science Institute, 3700 San Martin Dr, Baltimore, MD 21218, USA}
\affiliation{{William H. Miller III Department of Physics and Astronomy, Johns Hopkins University, Baltimore, MD 21218, USA}}



\begin{abstract}
Adding to the RESOLVE and ECO Gas in Galaxy Groups (G3) initiative, we examine possible drivers of group-integrated HI-to-halo mass ratios ($M_{\rm HI,grp}/M_{\rm halo}$) and group X-ray emission, including group halo mass ($M_{\rm halo}$), virialization as probed by crossing time ($t_{\rm cross}$), presence of active galactic nuclei (AGN), and group-integrated fractional stellar mass growth rate (FSMGR$_{\rm grp}$). G3 groups span $M_{\rm halo}=10^{11-14.5}\,M_\odot$ with comprehensive HI and AGN information, which we combine with X-ray stacking of ROSAT All-Sky data. We detect hot gas emission exceeding AGN and X-ray binary backgrounds confidently for $M_{\rm halo}=10^{12.6-14}\,M_\odot$ and unambiguously for $M_{\rm halo}>10^{14}\,M_\odot$, reflecting an inverse dependence of $M_{\rm\,HI,grp}/M_{\rm halo}$ and hot gas emission on halo mass. At fixed halo mass, $M_{\rm\,HI,grp}/M_{\rm halo}$ transitions to greater spread below $t_{\rm cross}\sim2$ Gyr. Dividing groups across this transition, lower-$t_{\rm cross}$ groups show elevated X-ray emission compared to higher-$t_{\rm cross}$ groups for $M_{\rm halo}>10^{13.3}\,M_\odot$, but this trend reverses for $M_{\rm halo}=10^{12.6-13.3}\,M_\odot$. Additionally, AGN-hosting halos below $M_{\rm halo}\sim10^{12.1}\,M_\odot$ exhibit a broad, $\sim$0.25 dex deep valley in $M_{\rm HI,grp}/M_{\rm halo}$ compared to non-AGN-hosting halos with correspondingly reduced FSMGR$_{\rm grp}$. When diluted by non-AGN-hosting halos, this valley becomes shallower and narrower, falling roughly between $M_{\rm halo}=10^{11.5}\,M_\odot$ and $M_{\rm halo}=10^{12.1}\,M_\odot$ in the overall $M_{\rm\,HI,grp}/M_{\rm\,halo}$ vs. $M_{\rm halo}$ relation. We may also detect a second, less easily interpreted valley at $M_{\rm halo}\sim10^{13}\,M_\odot$. Neither valley matches theoretical predictions of a deeper valley at or above $M_{\rm halo}=10^{12.1}\,M_\odot$.
\end{abstract}



\section{Introduction} \label{sec:intro}
Neutral atomic hydrogen (\HI{}) is the primary mass component of the interstellar medium in galaxies like the Milky Way \citep{kalberla2009hi}. Given its role in fueling star formation and AGN activity, \HI{} is a crucial ingredient in galaxy evolution. To investigate how environment affects the \HI{} content of galaxies and groups, recent observations and models have addressed the group \HI{}--halo mass relation (here, the $M_{\rm HI,grp}$--$M_{\rm halo}$ relation, where ``halo'' refers to the shared group halo; if a halo contains only one galaxy, we call it an $N_{\rm galaxies}=1$ group). The \relation{} describes the total \HI{} mass contained by a halo as a function of dark matter halo mass, and as such, its shape and scatter reflect cosmic gas accretion, assembly history, and feedback from active galactic nuclei (AGN) and star formation \citep{obuljen2019hi}. Previous $z\sim 0$ observations of this relation indicate (i) a transition to a shallower slope above $M_{\rm halo}\sim 10^{11.5}\, M_\odot$, corresponding to the peak integrated \HI{}-to-halo mass ratio, and (ii) a substantial scatter of at least $\sim$0.3 dex, suggesting that secondary factors regulate group \HI{} content at fixed halo mass \citep{eckert2017baryonic,obuljen2019hi,guo2020direct,dev2023galaxy, hutchens2023resolve, saraf2024xgass}. In this paper, we investigate some of these secondary factors --- virialization state, star formation, and AGN content --- and relate them to both \HI{} content and hot gas traced by X-ray emission.

In theoretical models, virial shocks are believed to be the primary halo mass-dependent mechanism of gas heating in groups. In the model of \citet[hereafter DB06]{dekel2006galaxy}, the onset of virial shock heating within $0.1R_{\rm vir}$ occurs at the same $M_{\rm halo}\sim 10^{11.5}\, M_\odot$ scale where the \relation{} transitions in slope. This ``gas-richness threshold scale'' (as named by \citealp{kannappan2009s0} due to the preponderance of \HI{} gas-dominated galaxies below it; see also \citealp{dekel1986origin, garnett2002luminosity, dalcanton2004formation, kannappan2004linking, kannappan2013connecting}) corresponds approximately to a stellar mass for the central galaxy of $M_*\sim 10^{9.5}\, M_\odot$ via the stellar mass--halo mass relation \citep{behroozi2010comprehensive}. Below the corresponding halo mass, theory predicts efficient gas cooling \citep{birnboim2003virial, kerevs2005galaxies, dekel2006galaxy, nelson2013moving}, as evidenced by the rapid refueling and stellar mass growth of galaxies in this regime \citep{kannappan2013connecting}. Slow, inefficient accretion is expected above the higher $M_{\rm halo} \sim 10^{12.1}\, M_\odot$ ``bimodality scale'' (corresponding to central galaxy $M_* \sim 10^{10.5 }\, M_\odot$), where the \citet{dekel2006galaxy} model predicts full halo gas heating (see also, e.g., \citealp{gabor2015hot}) and where observations show changes in galaxy morphology and stellar populations \citep{kauffmann2003dependence}. {These critical transitions in galaxy properties suggest that halo-driven gas heating may suppress galaxy \HI{} content via stripping and starvation. While hot gas halos have been detected in high-mass groups \citep{sun2009chandra, eckmiller2011testing, anderson2015unifying, jakobs2018multiwavelength} and down to $M_{\rm halo} = 10^{11}\, M_\odot$ \citep{kim2013scaling, goulding2016massive, forbes2017sluggs, bogdan2022x, zhang2024hot}, {these studies have generally not examined the interdependence of hot gas and \HI{} gas as a function of halo mass or other group properties, such as virialization state. Prior work connecting group hot and cold gas content has largely focused on single groups or small samples of compact groups (e.g., \citealp{rasmussen2012hot, desjardins2014some, o2018origin}).}

The dynamical state of a group {may affect} gas heating at fixed halo mass. As the dynamical evolution of a group progresses, simulations suggest that major and minor mergers may result in shocks that heat, and generate turbulence within, the intragroup medium \citep{sinha2009numerical, shi2020dynamical}. In a small sample of NGC groups, \citet{wilcots2009evolution} found that dynamically evolved groups host predominantly hot and ionized gas content, whereas dynamically young groups are more \HI{}-rich.   {The importance of group assembly in regulating \HI{} content is demonstrated in the results of \citet{hess2013evolution}, who found that \HI{}-rich galaxies preferentially reside on the outskirts of groups, and that the infall of \HI{}-rich satellites is crucial to replenish \HI{} in group halos.} Additionally, in a study of 172 SDSS groups with halo masses $M_{\rm halo} \gtrsim 10^{13}\, M_\odot$, \citet{ai2018evolution} showed that group \HI{}-to-halo mass ratios decrease with decreasing crossing time, corresponding to lower-$t_{\rm cross}$ states. Further analysis is needed to assess whether the relationship between group \HI{} content and crossing time extends to lower-mass halos, and if so, what physical processes drive the relationship.

Star formation (SF) and AGN are also expected to affect group \HI{} content at fixed halo mass, via feedback within halos. In semi-analytic models, \citet{eckert2017baryonic} found that ratios of hot halo gas to cold galaxy gas become widely varying in the $M_{\rm halo} \sim 10^{11.4-12.1}\,M_\odot$ regime, likely reflecting a transition in the dominant feedback source from SF to AGN. In lower-mass halos, SF feedback is expected from supernovae or the winds of young, massive stars; in higher-mass halos, AGN feedback is expected to heat or expel halo gas, thereby suppressing star formation or gas cooling \citep{somerville2008semi, gaspari2014can, fielding2017impact}. AGN feedback in dwarf galaxies may also lead to \HI{} suppression in low-mass halos, {as seen in the model of \citet{dashyan2018agn} and observations of \citet{bradford2018effect}, in which dwarf galaxies are defined by virial mass $M_{\rm vir}<10^{11}\, M_\odot$ and stellar mass $M_{\rm *}<10^{9.5}$, respectively.} Without special attention to dwarf AGN, recent theoretical models of the \relation{} have predicted an AGN-driven ``dip'' at $M_{\rm halo}\sim 10^{12.1-12.5}\, M_\odot$ \citep{kim2017spatial, baugh2019galaxy, chauhan2020physical}. This dip has evaded direct observation, possibly being eroded by observational systematics \citep{chauhan2021unveiling}. A fresh look with attention to dwarf AGN is warranted, given dramatic improvements in their detection \citep{polimera2022resolve, mezcua2024manga}.

In this paper, we ask three questions about the drivers of group \HI{} and hot gas:
\begin{enumerate}
    \itemsep0em
    \item How do group cold gas and X-ray emission depend on halo mass?
    \item How do group cold gas and X-ray emission depend on virialization state at fixed halo mass?
    \item How do group cold gas and X-ray emission depend on   {AGN prevalence and recent ($\sim$last Gyr) star formation history} at fixed halo mass?
\end{enumerate}
To answer these three questions, we combine archival \rosat{} All-Sky Survey data (RASS; \citealp{voges1993rosat}) with data from the Gas in Galaxy Groups (G3) initiative, a spinoff of the highly complete and volume-limited RESOLVE (REsolved Spectroscopy Of a Local VolumE; \citealp{kannappan2008galaxy}) and ECO (Environmental COntext; \citealp{moffett2015eco}) surveys. RESOLVE and ECO provide a complete census of a combined $\sim$456,300 Mpc$^{-3}$ volume of the $z\sim 0$ Universe, with comprehensive \HI{} mass, star formation history, and AGN data extending down to the dwarf regime, as needed to answer these questions. In G3 Paper I (\citealp{hutchens2023resolve}, hereafter H23), we constructed a group catalog for these surveys with optimal purity, completeness, and halo mass estimation. By stacking RASS imaging of G3 groups, we can analyze X-ray emission in relation to group \HI{} content, virialization state, star formation, and AGN activity.

This paper proceeds as follows. We describe the RASS data and the optical group catalogs for the G3 initiative in \S\ref{sec:data}. We describe our methods for reprocessing and stacking RASS imaging in \S\ref{sec:rassprocessing}. We outline our results in \S\ref{sec:results} and discuss their implications in \S\ref{sec:discussion}. Finally, we summarize our findings in \S\ref{sec:concl}. Throughout this work, we adopt a $\Lambda{\rm CDM}$ cosmology with $H_0 = 70\, {\rm km\, s^{-1}}\, \rm Mpc^{-1}$, $\Omega_{\rm m,0}=0.3$, and $\Omega_{\Lambda, 0}=0.7$.
\section{Data} \label{sec:data}

\subsection{RESOLVE and ECO Gas in Galaxy Groups}\label{subsec:data:g3}
In this section, we describe RESOLVE, ECO, and the G3 group catalogs.

\subsubsection{RESOLVE}\label{subsubsec:data:resolve}
RESOLVE is a highly complete, volume- and luminosity-limited census of stellar, gas, and dynamical mass in $\sim$53,000 Mpc$^3$ of the local Universe \citep{kannappan2008galaxy}. It contains $\sim$1,600 galaxies in two equatorial strips, RESOLVE-A ($131.25^{\rm o} < {\rm R.A.} < 236.25^{\rm o}$, $0^{\rm o} < {\rm Decl.} < 5^{\rm o}$) and RESOLVE-B ($330^{\rm o} < {\rm R.A.} < 45^{\rm o}$, $-1.25^{\rm o} < {\rm Decl.} < +1.25^{\rm o}$).\footnote{For both RESOLVE and ECO, our reported R.A. and Decl. ranges are in the J2000 coordinate frame.} Both RESOLVE-A and RESOLVE-B are limited by $4500 < cz_{\rm grp} < 7000$, where $cz_{\rm grp}$ is the Local Group-corrected recessional velocity of the galaxy's group, described further in \S\ref{subsubsec:data:g3groups}. RESOLVE-A is complete to $M_r = -17.33$ in the SDSS $r$ band, while RESOLVE-B (overlapping the deeper Stripe 82 region) is complete to $M_r = -17.0$ \citep{eckert2015resolve, eckert2016resolve}. Thus, defining dwarfs as galaxies with $M_r\geq -19.5$, RESOLVE is dwarf-dominated. RESOLVE data products include custom-reprocessed UV-to-NIR photometry as well as stellar masses and star formation history data derived from stellar population synthesis modeling (\citealp{eckert2015resolve}, based on \citealp{kannappan2013connecting}).

RESOLVE also provides a comprehensive atomic gas census that offers \HI{} detections or strong upper limits ($1.4M_{\rm HI}/M_* \lesssim 0.05-0.1$) for $\sim$94\% of galaxies (\citealp{stark2016resolve}, updated in H23). These measurements were derived from targeted Arecibo and Green Bank Telescope observations \citep{stark2016resolve} and supplemented by archival data from ALFALFA \citep{haynes2011arecibo, haynes2018arecibo} and \citet{springob2005digital}. Cases of confusion were identified automatically by searching for companions in existing redshift surveys, and deconfusion was attempted as outlined by \citet{stark2016resolve}. {A majority (81\%) of RESOLVE's \HI{} gas masses are based on such clean \HI{} detections, successfully deconfused detections, or strong upper limits, which we treat as best estimates of \HI{} mass. For the remaining 19\% of cases (missing 21cm observations, weak upper limits, or confused detections that could not be deconfused), H23 estimated \HI{} gas masses using the photometric gas fractions technique \citep{kannappan2004linking, eckert2015resolve}, as constrained by weak upper limits or confused total fluxes when available.} 

{In addition to photometry and \HI{} mass information, the present work uses RESOLVE's AGN inventory (M.S. Polimera et al. 2025, in prep.). One of the advantages of our analysis is that this AGN catalog provides a uniform and unusually complete selection of low-mass galaxy AGN, found using both optical emission line diagnostics (which reveal the new ``SF-AGN'' class of low-metallicity/star-forming AGN found by \citealp{polimera2022resolve} and additional dwarf AGN identified by the loosened BPT diagnostic of \citealp{stasinska2006semi}) and mid-IR color diagnostics (which are sensitive to a complementary set of dwarf AGN candidates; \citealp{satyapal2018star}; M.S. Polimera et al. in prep., 2025; see \S 4.3 regarding reliability concerns for mid-IR AGN). Additionally, the catalog provides cross-matched data on previously identified X-ray AGN \citep{ranalli2003xray}, broad-line AGN \citep{liu2019comprehensive}, and other AGN cross-matched from the literature \citep{veroncetty2006catalogue, flesch2015half}.}

\subsubsection{ECO}\label{subsubsec:data:eco}
The Environmental COntext (ECO; \citealp{moffett2015eco}, updated by H23) catalog surrounds RESOLVE-A in a $\sim$10 times larger volume of $\sim$440,000 Mpc$^{-3}$. {ECO's larger volume supports the smaller but superior RESOLVE survey by enabling larger-sample studies of environment and assessment of cosmic variance}. The ECO volume is defined by $3000 < cz_{\rm grp} < 7000$, $130.05^{\rm o} < {\rm R.A.} < 237.45^{\rm o}$, and $-1^{\rm o} < {\rm Decl.} < 49.85^{\rm o}$. ECO reaches the same luminosity completeness floor as RESOLVE-A ($M_r = -17.33$), but is purely archival, except where new observations were incorporated via its overlap with RESOLVE-A. All ECO data products, including photometry and \HI{} mass information, were processed using RESOLVE pipelines to improve quality and harmonize the two surveys. {Like RESOLVE, ECO offers a uniform and unusually complete AGN catalog based on multiple optical emission line diagnostics and mid-IR colors, which also provides cross-matched data on known broad-line AGN, X-ray AGN, and other AGN from the literature (see \S\ref{subsubsec:data:resolve}).}

{Additionally, ECO contains a flux-limited census of \HI{} gas comprised of inherited RESOLVE-A observations and cross-matched sources from ALFALFA \citep{haynes2018arecibo}, including new upper limits and confusion flags (see H23). As for RESOLVE, H23 computed the best \HI{} mass estimate for each ECO galaxy by combining clean \HI{} detections, upper limits, confused fluxes, and constrained photometric gas fraction estimates (see \S\ref{subsubsec:data:resolve}), with the exception that deconfusion was not attempted for confused ECO galaxies outside RESOLVE-A. For these confused sources in ECO, \HI{} mass estimates are always from photometric gas fractions constrained by confused fluxes. For the entire ECO survey, 45\% of our best \HI{} mass estimates are based on clean detections, strong upper limits, or (in the RESOLVE-A overlap region only) de-confused observations; the remaining 55\% are photometric gas fraction estimates for galaxies with weak upper limits, confused 21cm observations, or missing 21cm observations.}

\subsubsection{G3 Groups}\label{subsubsec:data:g3groups}
The group catalogs used for our X-ray stacking analysis were created by H23 using the Gas in Galaxy Groups (G3) group finder, whose four-step algorithm offers improved completeness and halo mass recovery compared to friends-of-friends group finding. 
For the analysis in this paper, we combine the G3 group catalogs from RESOLVE-B and ECO (where the latter includes RESOLVE-A). We selected groups whose most luminous galaxies (which we refer to as ``centrals'') are above the survey luminosity floors and whose average group redshifts fall within the respective survey redshift ranges of $3000 < cz_{\rm grp}{\,\left[{\rm km\, s^{-1}}\right]\,} < 7000$ (ECO) or $4500 < cz_{\rm grp} {\,\left[{\rm km\, s^{-1}}\right]\,} < 7000$ (RESOLVE-B). The selection on $cz_{\rm grp}$ mitigates the clipping of groups when galaxy peculiar velocities would otherwise have extended beyond the survey redshift limits. The resulting selection yields 6,949 groups with halo masses spanning $M_{\rm halo} = 10^{11}-10^{14.5}\, M_\odot$, consisting of 6,038 $N_{\rm galaxies}=1$ groups, 512 galaxy pairs, and 399 groups with $\geq3$ galaxies. We exclude the Coma cluster from our stacking analyses, as its intense X-ray emission makes it an outlier among our other massive groups, so our final sample consists of 910 groups/pairs and 6,038 $N_{\rm galaxies}=1$ groups.

\subsubsection{G3 Group Properties}
In our analysis in \S\ref{sec:results}, we use RESOLVE and ECO data to compute four main group properties needed to answer the questions introduced in \S\ref{sec:intro}:
\begin{enumerate}
    \item Group halo mass $M_{\rm halo}$: Group halo masses for G3 groups were derived in H23 using abundance matching \citep{kravtsov2004dark, blanton2007aspects}. With this technique, H23 built a one-to-one monotonic relationship between group halo mass and group-integrated $r$-band luminosity using the theoretical halo mass function of \citet{tinker2008toward}. {The derived halo masses and radii assume a mean background overdensity of $\Delta_{\rm vir} = 337$, representing the boundary of the virialized halo (e.g., \citealp{d2005formation}). Compared to the common alternatives of 200 or 500, our halo masses scale as $M_{337} = 0.85M_{200}$ and $M_{337}=1.1M_{500}$. Since halo radii $R_{\rm halo}\propto M_{\rm halo}^{1/3}$, our virial radii scale as $R_{337} = 0.95R_{200}$ and $R_{337}=1.04R_{\rm 500}$ (see \citealp{hu2003sample}).}
    
    \item Group-integrated \HI{} Mass $M_{\rm HI,grp}$: As in H23, we compute group-integrated \HI{} mass by summing galactic \HI{} mass estimates for group members.\footnote{{Given the sizes of the Arecibo and GBT beams, our galactic \HI{} masses may sometimes include intragroup gas and/or gas-rich, optically undetected satellite galaxies. However, given that intragroup \HI{} outweighs galactic \HI{} by at most a factor of 1.5--2 even in intragroup-dominant systems \citep{van1992study, borthakur2015distribution}, we do not expect a significant contribution from intragroup \HI{} affecting our results. In fact, our group \HI{} masses are largely consistent with those of \citet{guo2020direct}, who measured the \HI{}--halo mass relation using a stacking method that intrinsically includes intragroup \HI{} (see H23).}} These \HI{} mass estimates are a combined dataset of clean \HI{} detections, strong upper limits, deconfused \HI{} detections, and estimates {from photometric gas fractions (possibly constrained by upper limit/confused flux data; see H23 for a description of how this method incorporates upper limit and confused data).} For $N_{\rm galaxies}=1$ groups, $M_{\rm HI,grp}$ is the ${\rm galaxy\,} M_{\rm HI}$. In \S\ref{sec:results}, we derive errors on median $M_{\rm HI,grp}/M_{\rm halo}$ values using bootstrapping with 5,000 resamples.

    \item Group-integrated fractional stellar mass growth rate $\rm FSMGR_{\rm grp}$: The galaxy fractional stellar mass growth rate (${\rm FSMGR}$) is the ratio of stellar mass formed within the past Gyr, $M_{*, \rm 1\,Gyr}$, to the preexisting stellar mass formed over all previous Gyr, $M_{*, \rm preex}$ \citep{kannappan2013connecting}.   {We calculate FSMGR using the SED modeling code of \citet[based on \citealp{kannappan2007systematic}]{kannappan2013connecting}, which uses our custom NUV to IR photometry (see \citealp{eckert2015resolve, moffett2015eco}; H23).} {We prefer FSMGR as a SF metric over the common alternative, specific star formation rate (${\rm sSFR} = {\rm SFR}/M_*$), because FSMGR is better suited to probe high fractional growth regimes of star formation: an FSMGR can reach arbitrarily large values, whereas an sSFR approaches an asymptotic maximum as recent stellar mass growth increases, contributing to both SFR and $M_*$.} We calculate group-integrated FSMGR as ${\rm FSMGR_{\rm grp}} = (\sum M_{*,\rm 1\,Gyr})/(\sum M_{*,\rm\, preex})$, where these sums are computed over all group members. For $N_{\rm galaxies}=1$ groups, $\rm FSMGR_{\rm grp}$ is the galaxy FSMGR. In \S\ref{sec:results}, we derive errors on median \fsmgrgrp{} values using bootstrapping with 5,000 resamples.
    
    \item Group crossing time $t_{\rm cross}$: To assess virialization state, we use the crossing time $ t_{\rm cross} = {\left<R_{\rm proj,gal}\right>}/{\left<|v_{\rm proj,gal}|\right>}$
    (\citealp{rood1978virial}; see also \citealp{ferguson1990population,firth2006kinematics}), where $\left<R_{\rm proj,gal}\right>$ and $\left<|v_{\rm proj,gal}|\right>$ are the mean projected transverse distance and mean absolute line-of-sight velocity, respectively, for grouped galaxies relative to their average group center (see \S\ref{subsec:data:rass} below). This metric represents the mean time for a galaxy to traverse the group, and is usually expressed relative to the age of the Universe \citep{hickson1992dynamical}. {A system of galaxies freely expanding with the Hubble flow will have a crossing time comparable to the age of the Universe, whereas if the crossing time is short compared to the age of the Universe, then the group is a bound, virialized system \citep{gott1977groups}.} Unlike other virialization metrics (e.g., the Dressler-Shectman statistic; \citealp{dressler1988evidence}), $t_{\rm cross}$ can be calculated for all $N_{\rm galaxies}\geq 2$ groups. {However, we note that $t_{\rm cross}$ is a noisy metric of virialization, being both subject to observational projection effects and sensitive to abrupt changes due to merging (e.g., when the smallest $R_{\rm proj,\,gal}$  values are eliminated after a merger).}
\end{enumerate}

\subsection{ROSAT All-Sky Survey Data}\label{subsec:data:rass}
We use archival X-ray imaging from the \rosat{} All-Sky Survey (RASS; \citealp{voges1993rosat,snowden1994analysis}). The RASS survey was conducted in scanning mode using \rosat{}'s Position Sensitive Proportional Counters (PSPC) to observe 1,378 $6.4^{\rm o} \times 6.4^{\rm o}$ fields in three energy bands. 
We obtained broad band ($\sim$0.1--2.3 keV, PSPC channels 11--235) and hard band ($\sim$0.44--2.04 keV, PSPC channels 52--201) photon count maps, exposure maps, and background count maps via HEASARC\footnote{https://heasarc.gsfc.nasa.gov/} (High Energy Astrophysics Science Research Center). Since we obtained the most significant results in the RASS hard band, we do not show broad band stacking results in this paper. For each G3 group, we extracted custom-mosaicked maps at the average group center (see H23), using the Python library \texttt{reproject} to ensure flux conservation. While the location of the brightest cluster galaxy (BCG) is a common alternative to the average group center, \citet{skibba2011brightest} showed that 25\%--40\% of BCGs in halos of mass $M_{\rm halo}>10^{12.1}\,M_\odot$ are not actually the galaxies with the lowest specific potential energies in their halos. Moreover, \citet{gozaliasl2019chandra} measured considerable 0.2--0.3$R_{\rm vir}$ offsets between BCGs and X-ray centroids of groups, which depend on both halo mass and magnitude gap (probing virialization state). These findings suggest that the position of the BCG is not the optimal location to stack X-ray emission in groups.


\section{RASS Image Processing \& Stacking} \label{sec:rassprocessing}
In this section, we outline our four-step strategy for reprocessing and stacking RASS data: (1) image masking to remove extraneous point and diffuse sources, (2) image scaling and cropping to enable stacking on a common scale, (3) group image stacking, and (4) random image stacking.

\begin{figure*}
    \centering
    \includegraphics[scale=0.85]{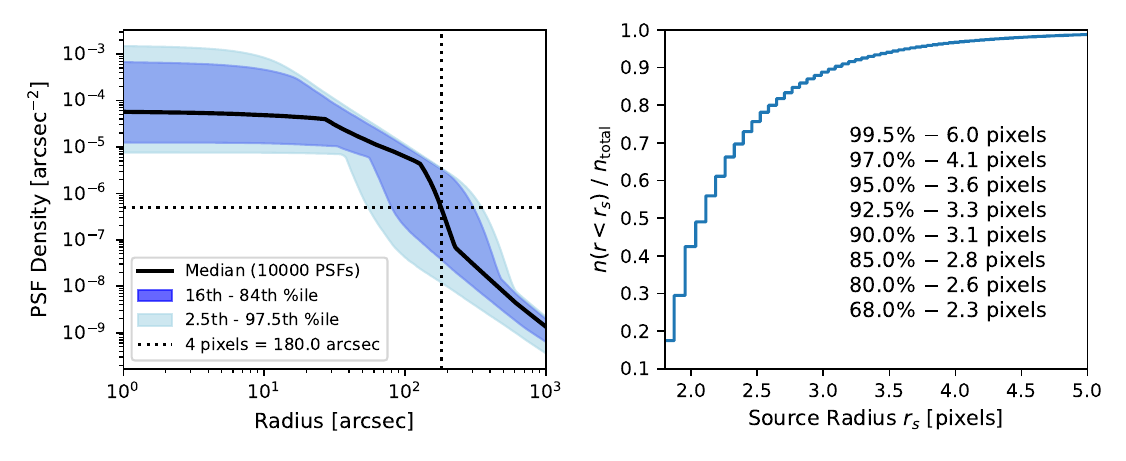}
    \caption{PSPC PSF and distribution of source radii. {Left:} Distribution of 10,000 \textsl{ROSAT} PSPC PSFs calculated with random photon energies spanning 0.44--2.04~keV and random off-axis angles spanning $0^{\prime}$--$60^{\prime}$. The solid black line shows the median PSF; blue shaded regions show the 16th--84th (dark) and 2.5th--97.5th (light) percentiles of the distribution at fixed radius. Dotted lines show the PSF strength at a radius of 4 pixels (180$^{\prime\prime}$). {Right: } Cumulative distribution function $n(r < r_s)/n_{\rm total}$ describing the fraction of \textsl{ROSAT} sources, as identified with the iterative sigma-clipping algorithm described in \S\ref{subsec:masking}, with radii less than $r_s$. Annotated values tabulate the percentage of sources with radii less than $r_s$.
    }
    \label{fig:mask_size}
\end{figure*}

\subsection{Source Masking}\label{subsec:masking}
We implemented source masking to exclude nearby extraneous sources as well as to separate point-source (including unresolved galaxy) emission from extended emission within our groups. To do so, we applied an iterative sigma-clipping algorithm from the \texttt{photutils} Python library \citep{price2022astropy} to detect and then mask any $\geq 5\sigma$ sources lying outside the group virial radius in each image. This code creates a segmentation map in which sources may have arbitrary geometries, allowing us to mask both point and extended sources. In addition, we masked sources listed in the Second \rosat{} X-ray Source catalog (2RXS; \citealp{boller2016second}). To select sources for masking from 2RXS, we excluded 2RXS objects with detection likelihood \texttt{EXI\_ML} $\leq 9$, which lowers the 2RXS spurious source fraction to $\sim$5\% (see \citealp{boller2016second}).   {We also excluded nine 2RXS sources that cross-matched to RESOLVE or ECO galaxies within $6^{\prime\prime}$ (the median effective radius of RESOLVE/ECO galaxies), as for some analyses we choose to include galactic emission and do not want these galaxies to be automatically masked.} We automatically masked the remaining 2RXS point sources {(\texttt{EXI\_ML} $>9$ and not cross-matched to RESOLVE/ECO)} as described below. 

To determine mask apertures for both RESOLVE/ECO AGN galaxies and 2RXS sources, we considered whether these sources could be matched to sources identified by our sigma-clipping algorithm. If so, we extracted masks from the segmentation map, which naturally provides a custom aperture for each source. If the RESOLVE/ECO AGN or 2RXS source was not detected by the sigma-clipping algorithm, we applied a circular mask with radius 4 pixels, where each RASS pixel is $45^{\prime\prime}$. As illustrated in Figure \ref{fig:mask_size}, this value was chosen because $\sim$97\% of the sources identified by the sigma-clipping algorithm can be enclosed by such an aperture. Figure \ref{fig:mask_size} also shows that the median PSF of the PSPC reaches $\sim$1\% of its maximum strength at this 4-pixel radius. We compute the median PSF from a sample of PSFs generated at random 0.44--2.04 keV photon energies and $0^{\prime}-60^{\prime}$ off-axis angles (the PSPC field of view), based on the analytic formula for the PSPC PSF \citep{boese2000rosat}. This statistical PSF computation is necessary because RASS imaging data are coadds of multiple visits to each field; thus, sources in the RASS data have been observed at a variety of unknown off-axis angles.

In \S\ref{sec:results}, to separate galactic and intragroup X-ray emission, we provide stacking results both with AGN masked and with AGN unmasked. In the former, we masked the locations of all RESOLVE and ECO AGN contained within the custom-mosaicked count maps. For both masked AGN and 2RXS sources, we replace masked regions with the background count level. Figure \ref{fig:masking_coma} illustrates an example of this masking procedure.

\begin{figure*}
    \centering
    \includegraphics[scale=0.9]{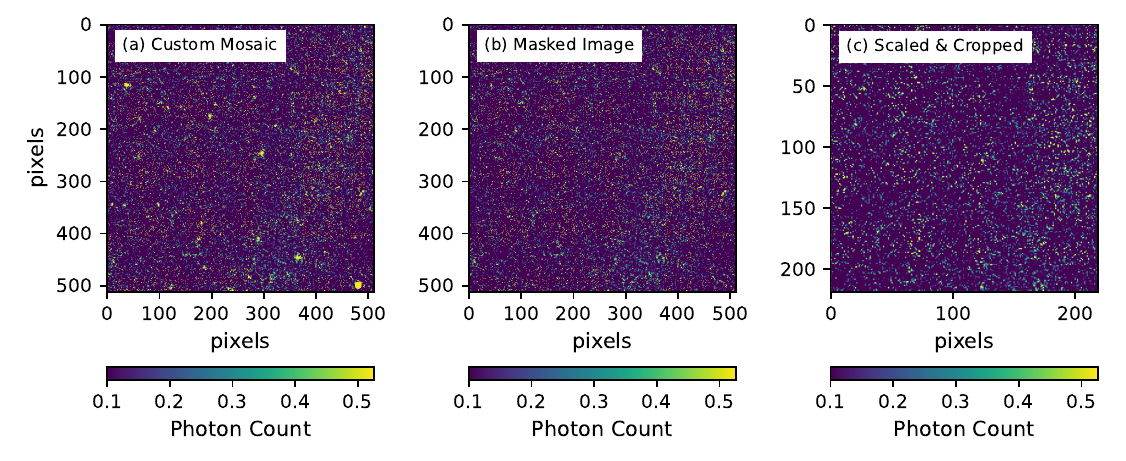}
    \caption{Demonstration of RASS image mosaicking, masking, and scaling/cropping applied to G3 group \#93 (chosen arbitrarily for illustration). (a) Custom count map centered on group \#93. Pixel values are in units of photon counts and may be fractional given that the original RASS data have been reprojected and mosaicked as described in \S\ref{subsec:data:rass}. (b) Masked count map, with 2RXS point sources masked, as outlined in \S\ref{subsec:masking}. (c) Scaled image cropped to a $219 \times 219$ pixel cutout, as detailed in \S\ref{subsec:scaling}.}
    \label{fig:masking_coma}
\end{figure*}

\subsection{Scaling and Cropping Images}\label{subsec:scaling}
Since we wish to stack images of groups at different distances, it is crucial that we add RASS images on a consistent scale. Following \citet{dai2007x}, we scaled our masked group images to a common kpc/degree scale matching the outer distance of the ECO survey (100 Mpc for our chosen Hubble constant $H_0 = 70\,\rm km\,s^{-1}Mpc^{-1}$) to ensure that all pixel resampling combines rather than oversamples pixels. The original $512 \times 512$ masked mosaics are then contained within an inner square region of the scaled image, bordered by zeros. The scaled images conserve the number of photons in the original image, with a typical error of $<$1\%. Conservation of photon number is appropriate, as this rescaling procedure is not meant to simulate artificial redshifting of groups. Rather, it enables us to stack individual group images on a common physical scale, i.e., to ensure that halos of the same mass are stacked consistently even if they are located at different distances. We also applied this same scaling procedure to our exposure and background count maps, which are needed to perform stacking.

Prior to stacking, we cropped our scaled count, exposure, and background maps to centered $219 \times 219$ pixel cutouts. At the RASS $45^{\prime\prime}$ pixel scale and our outer survey distance limit of 100 Mpc, the 219 pixel image width corresponds to 4.7 Mpc on-sky. Importantly, cropping to $219 \times 219$ removes the zero-containing border regions left over from image scaling, even for the nearest ECO galaxy groups at $\sim$42.9 Mpc (noting that $512\,{\rm pixels}\times 42.9{\,\rm Mpc} / 100 {\,\rm Mpc} = 219 {\,\rm pixels}$), but still comfortably contains our largest groups with $R_{\rm vir} \sim 1.7$ Mpc.  The last panel of Figure \ref{fig:masking_coma} shows an example of a scaled and cropped count map.

\subsection{Image Stacking and SNR Calculation}\label{subsec:stacking_snr}
In \S\ref{sec:results}, we present X-ray stacking results binned by halo mass and other group properties. To construct the stacked images, we applied stacking methods from prior RASS analyses  \citep{dai2007x, anderson2012extended, anderson2015unifying}.  For each stacking bin consisting of a set of $n_{\rm bin}$ groups, we have a set of masked count maps $\{C_1, C_2, ..., C_{n_{\rm bin}}\}$, exposure maps $\{E_1, E_2, ..., E_{n_{\rm bin}}\}$, and background maps $\{B_1, B_2, ..., B_{n_{\rm bin}}\}$. We summed these sets to obtain a stacked count map $\mathcal{C} = \sum_{i=1}^{n_{\rm bin}} C_i$ in counts, a stacked exposure map $\mathcal{E} = \sum_{i=1}^{n_{\rm bin}} E_i$ in seconds, and a stacked background map $\mathcal{B} = \sum_{i=1}^{n_{\rm bin}} B_i$ in counts. We compute the final background-subtracted intensity map for each stack as $\mathcal{R} = (\mathcal{C}-\mathcal{B})/\mathcal{E}$, in units of counts per second.\footnote{We note that averaging intensity maps directly, i.e., $\mathcal{R} = (1/n_{\rm bin})\sum{(C_i-B_i)/E_i}$, will yield the same result when $E_i$ is the same for all groups. However, RASS coadd exposure times vary substantially across the sky (e.g., \citealp{voges1999rosat} Figure 1), so we follow \citet{dai2007x} and \citet{anderson2015unifying} in using $\mathcal{R} = (\mathcal{C}-\mathcal{B})/\mathcal{E}$. This definition provides a more robust average, especially for our low $n_{\rm bin}$ stacks.} 

{For each stack, we have accounted for source confusion by excluding pairs of groups whose angular separation is less than either of their individual virial radii (i.e., at least one of the groups resides within $R_{\rm vir}$ of the other in projection). However, since intragroup X-ray emission is typically concentrated to only 0.1--0.5$R_{\rm vir}$ \citep{mulchaey2000x}, and because low-mass groups are less likely to show bright X-ray emission, we made an exception if the two confused groups have drastically different mass ($>$1 dex) and are separated by more than $R_{\rm vir,\, min} + (1/2)R_{\rm vir,\,max}$, where $R_{\rm vir,\, min}$ is the virial radius of the lower-mass group and $R_{\rm vir,\, max}$ is the virial radius of the higher-mass group. In these cases, the lower-mass group is unlikely to contaminate the X-ray stack containing the higher-mass group, so we only excluded the lower-mass group. Based on this definition of confusion, we have flagged and excluded $\sim$21\% of our groups from stacking. The fraction of groups retained as a function of halo mass is approximately constant at $\sim$70-80\%}.

After constructing stacks and excluding confused objects, we computed the X-ray signal-to-noise ratio (SNR) within $0.5R_{\rm vir}$ for each stack, where $R_{\rm vir}$ is calculated using the halo mass at the bin center. The choice of $0.5R_{\rm vir}$ stems from the typical concentration of emission in X-ray--detected galaxy groups \citep{mulchaey2000x}. Following \citet{de1997steepness}, we calculate the SNR as $N_{\rm src}/\sqrt{N_{\rm total}}$, where $N_{\rm src}$ and $N_{\rm total}$ are the numbers of source counts and total (background + source) counts, respectively, within the $0.5R_{\rm vir}$ aperture. {We report an SNR of zero when the stacked background-subtracted count rate is zero or negative, which is possible because the RASS background maps are smooth whereas the RASS count maps are noisy.} {We also compute the total count rate within $0.5R_{\rm vir}$, as well as its uncertainty $\sigma_{\rm RMS}\sqrt{N}$, where $\sigma_{\rm RMS}$ is the root-mean-square noise of the intensity map and $N$ is the number of pixels in the $0.5R_{\rm vir}$ aperture.}

\subsection{Randomized Stacking}\label{subsec:randomizedstacking}
SNR alone is not sufficient to justify a claim that we have detected \emph{diffuse} X-ray emission associated with G3 groups, as (compact or diffuse) foreground or background X-ray sources landing within the group virial radius could create a false signal. To quantify possible contributions from such sources, we replicated our data processing and stacking procedures using a set of images extracted at random sky positions. We generated 4,093 (the total number of $N_{\rm galaxies}\geq 1$ groups with $M_{\rm halo} = 10^{11-11.5}\,M_\odot$, our largest stacking bin) sets of count maps, exposure maps, and background count maps at R.A. and Dec sampled randomly from the RESOLVE and ECO footprints, avoiding sky positions within projected $2R_{\rm vir}$ of any known groups. By extracting these images within the RESOLVE and ECO footprints, we have avoided Galactic X-ray emission at low Galactic latitudes and ensured that our random images have the same statistical properties as the G3 group images.

For each stacking bin containing $n_{\rm bin}$ G3 groups, we sampled the 4,093 image-sets with replacement to create {30} random image set samples. For each of the {30} samples, we masked, scaled, cropped, and stacked the random $n_{\rm bin}$ images so that they were processed identically to the image-sets for the corresponding $n_{\rm bin}$ G3 groups assigned to that bin. The final product for each bin is a set of {30} stacks, which allows us to analyze the distributions of SNR and integrated count rate found in random image stacks.

\subsection{Estimating X-ray Binary Contributions}\label{subsec:xrbmethod}
{In \S4, to help understand the origins of our stacked X-ray emission, we have provided for each stacking result the expected count rate from low-mass and high-mass X-ray binaries (LMXBs and HMXBs). To estimate these contributions, we considered multiple estimates of XRB scaling relations from the literature, including estimates for LMXB-only emission \citep{david2006hot, boroson2011revisiting, lehmer2020x}, HMXB-only emission \citep{mineo2012x, mineo2014xray, lehmer2022elevated}, and total XRB emission \citep{colbert2004old, lehmer2010chandra, lehmer2019x, lehmer2024empirical}. Including all $3\times3$ possible pairs of these LMXB-only and HMXB-only estimators, we obtained 13 distinct estimators of total XRB emission. Next, for each stack, we computed exposure-weighted mean group-integrated properties (e.g., SFR, sSFR, $L_{\rm K}$, $M_*$) as needed to evaluate these estimators and derive expected XRB luminosity estimates for the stack. Finally, we converted these luminosities to fluxes, and converted the fluxes to expected RASS count rates using the PIMMS software (Portable, Interactive, Multi-Mission Software; \citealp{mukai1993pimms}). We used each paper’s reported spectral model to convert to the PSPC hard band, correcting for foreground absorption using Galactic \HI{} column densities from the \texttt{gdpyc} Python library, which are based on the Leiden/Argentine/Bonn Galactic \HI{} Survey \citep{kalberla2005leiden}.} 

{
Using these 13 estimators, we report for each stack in \S\ref{sec:results} the minimum and maximum estimated XRB contributions and our preferred estimated XRB contribution. The XRB contribution increases with group halo mass for all estimators due to the underlying increase of group $N_{\rm galaxies}$ with increasing halo mass. We have chosen the relation of \citet{lehmer2019x} as our default XRB estimator because (i) it is derived using a robust statistical sample of nearby galaxies with diverse morphologies, masses, and star formation rates, and (ii) it predicts RASS count rates consistent with our measured AGN-masked count rates within the uncertainties at low $M_{\rm halo}$, where X-ray emission should be dominated by XRBs.\footnote{{We prefer the relation of \citet{lehmer2019x} to that of \citet{lehmer2024empirical} because the latter yields XRB estimates that are systematically higher than our measured count rates in bins where the emission should be dominated by XRBs. The estimates from \citet{lehmer2024empirical} exceed our AGN-masked count rates in the three halo mass bins spanning $M_{\rm halo} = 10^{11-12.6}\, M_\odot$, and they also exceed our total count rates (with AGN still included) in the $M_{\rm halo} = 10^{11-11.5}\, M_\odot$ and $M_{\rm halo}=10^{12.1-12.6}\,M_\odot$ bins (see \S4).}} Notwithstanding the good reasons for this choice of preferred estimator, we note that accurately estimating the contribution from XRBs is one of the greatest areas of uncertainty in this work and similar group X-ray stacking studies.}

{When analyzing X-ray stacks with AGN-host galaxies masked, we excluded these AGN-host galaxies' XRB contributions from our expected XRB count rate calculation,  as any XRB emission from these galaxies will have been excluded by the AGN masks.}

\section{Results}\label{sec:results}

\subsection{How do cold gas and group X-ray emission depend on halo mass?}\label{subsec:logmhstacking}

Figure \ref{fig:hardband_all_stack} shows stacked X-ray emission in six bins of group halo mass, chosen to coincide with regimes around the threshold and bimodality scales (see \S\ref{sec:intro}) and also to facilitate comparison with past work (e.g., \citealp{anderson2015unifying}). We provide total stacked count rates, stacked count rates with AGN galaxies masked, and expected XRB count rates. In all three cases, count rates increase as a function of halo mass. 

With galaxies and AGN not masked (``all X-ray emission''), we detect X-ray emission with SNR$>$2 in excess of the random stacks for all halo mass bins except  $M_{\rm halo} = 10^{11-11.5}\, M_\odot$, with more confident SNR$>$4 detections at $M_{\rm halo} = 10^{11.5-12.1}\, M_\odot$ and in the three individual bins spanning $M_{\rm halo} = 10^{12.6-14.5}\, M_\odot$. \footnote{The detection in the $M_{\rm halo} = 10^{12.6-13.3}\, M_\odot$ bin drops to {SNR$\sim$5.5 if we exclude group \#1345}, which contains an extremely X-ray-bright galaxy (ECO04631 / 2RXS J141759.5+250817).\label{footnote:ECO04631}} Interestingly, the $M_{\rm halo} = 10^{12.1-12.6}\, M_\odot$ bin shows a marginal 2.4$\sigma$ detection despite being surrounded by more significant detections in adjacent bins. We speculate that this result may represent the confluence of decreasing $n_{\rm bin}$ (resulting in {lower stacking depth}) and increasing intrinsic X-ray emission as halo mass increases. 

After AGN are masked, all count rates decrease relative to total count rates, albeit total and AGN-masked count rates are similar when considering their uncertainties. {With AGN masked, we detect X-ray emission in excess of random stacking expectations in the three bins spanning $M_{\rm halo} = 10^{12.6-14.5}\, M_\odot$, with the middle bin at $10^{13.3-14}\, M_\odot$ reaching only a marginal SNR of 2.7 and the flanking bins reaching SNRs of 3.5 and 5.4.}

Since masking AGN removes any AGN X-ray emission\footnote{Masking AGN may also mask intragroup hot gas in the foreground or background of the AGN-host galaxies, so our AGN-masked count rates are subject to this caveat. However, in each halo mass bin, the ratio of AGN-masked pixels to total stacked pixels is only $\sim$10\%.}, any remaining X-ray emission would be most likely attributed to XRBs or hot gas; thus, we compare to the expected X-ray binary contribution for each stack. {In the highest mass bin at $M_{\rm halo} = 10^{14-14.5}\, M_\odot$, our preferred XRB estimate accounts for only $6\pm1\%$ of the observed AGN-masked count rate. In the next lower bins at $M_{\rm halo} = 10^{13.3-14}\, M_\odot$ and $M_{\rm halo} = 10^{12.6-13.3}\, M_{\odot}$, our preferred XRB estimates account for only $24\pm5\%$ and $33\pm 4$, respectively, of the observed AGN-masked count rates. However, for these bins, our measured AGN-masked count rates are smaller than the highest estimates from the 13 XRB estimators we reviewed. Consequently, our results unambiguously confirm the presence of hot gas only for $M_{\rm halo} > 10^{14}\, M_\odot$, though we consider our hot gas detections in the two bins spanning $M_{\rm halo} = 10^{12.6-14}\, M_\odot$ to be reasonably secure. In lower halo mass bins, our observed AGN-masked count rates are consistent with or smaller than expectations for XRBs.}  

{For comparison with the X-ray count rates, the lower right panel of Figure \ref{fig:hardband_all_stack} shows the distribution of $M_{\rm HI,grp}/M_{\rm halo}$ as a function of halo mass. From the lowest to highest halo mass bin, the median $M_{\rm HI,grp}/M_{\rm halo}$ drops by 1.1 dex (a factor of $\sim$13) as the stacked X-ray intensity increases.} 

\begin{figure*}
    \centering
    \includegraphics[scale=0.13]{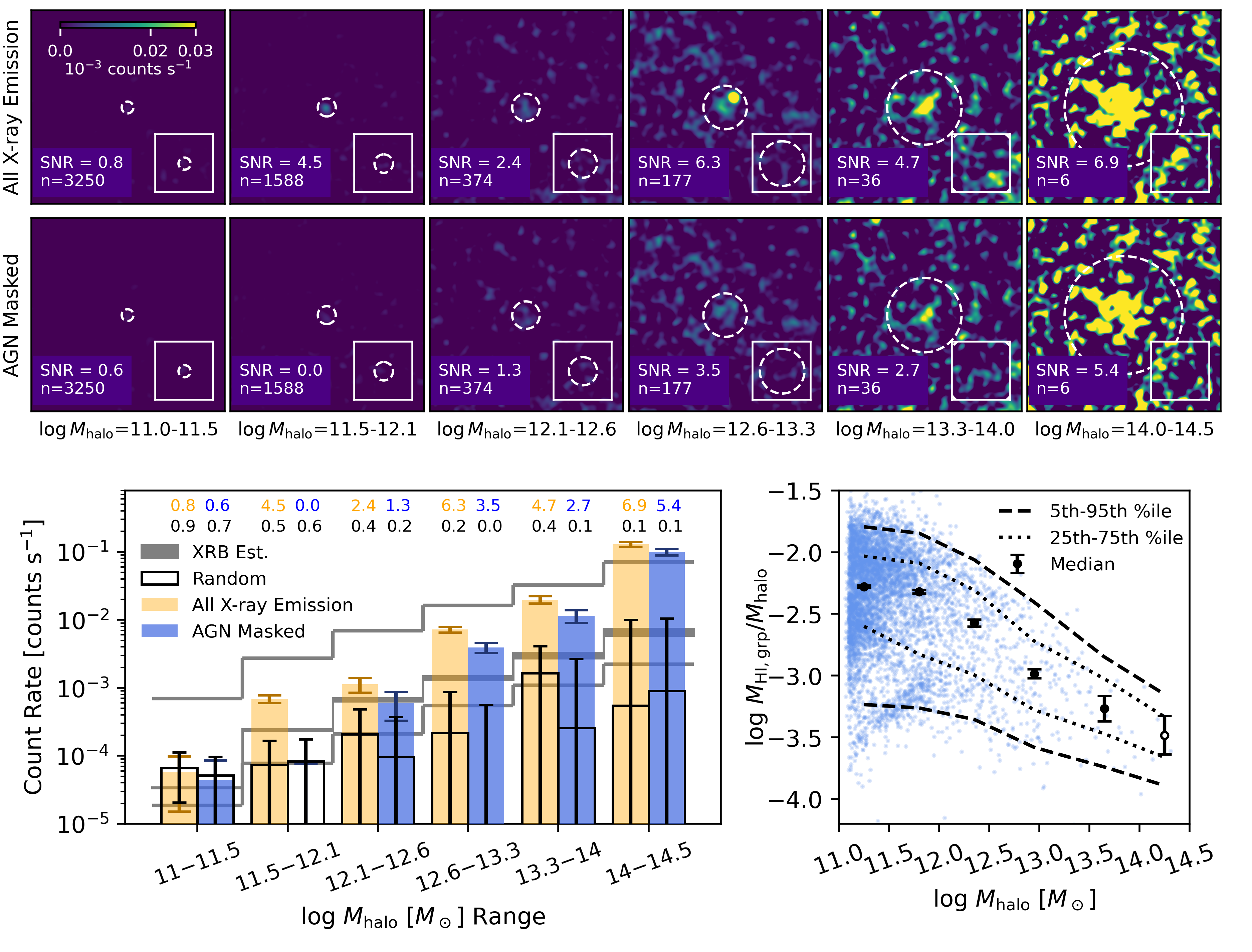}
    \caption{Stacked X-ray emission and group $M_{\rm HI,grp}/M_{\rm halo}$ in bins of group halo mass. $N_{\rm galaxies}=1$ groups are included. {Top: } Stacked X-ray intensity maps when AGN galaxies are not masked. Images are smoothed with a 3-pixel Gaussian kernel to enable visualization on a common linear scale, as logarithmic scales emphasize noise fluctuations in low-$n_{\rm bin}$ stacks. White circles represent $R_{\rm vir}$ for a group at the center of the halo mass bin,   {with values of 0.14, 0.22, 0.33, 0.54, 0.91, and 1.45 Mpc from left to right}. The annotations note the SNR and $n_{\rm bin}$ for each stack. {Middle: } Same as top row, but now with AGN masked. {Bottom Left:} Stacked, background-subtracted X-ray count rate as a function of halo mass for all intensity maps shown. Orange bars represent total count rates (AGN not masked); blue bars represent count rates with AGN masked. Black bars show the median count rate based on random stacking, and {gray lines show estimates for XRB contributions.  The top and bottom thin gray lines show the maximum and minimum estimates from the 13 different XRB estimators we considered (see \S\ref{subsec:xrbmethod}). The middle thick gray line shows our preferred estimates using the \citet{lehmer2019x} calibration, where the width of the line indicates the uncertainty in this estimator based on bootstrapping.} The annotations list the SNR for each stack, with SNRs from randomized stacks in black. {Bottom Right:} $M_{\rm HI,grp}/M_{\rm halo}$ vs. $M_{\rm halo}$ as in H23. Black points represent median values with errors determined using bootstrapping; open circles represent median values in bins with fewer than 30 groups, for which bootstrap errors may be unreliable. The dashed and dotted lines represent the 5th--95th and 25th--75th percentiles, respectively. }

    \label{fig:hardband_all_stack}
\end{figure*}

To highlight X-ray contributions driven by the group environment, Figure \ref{fig:cr_vs_mhalo_halomassbins} assesses stacked X-ray emission excluding $N_{\rm galaxies}=1$ groups and considering only galaxy groups with halo masses $M_{\rm halo} = 10^{11.5 - 14.5}\, M_\odot$, below which there are insufficient $N_{\rm galaxies}=1$ halos for stacking. With AGN not masked, Figure \ref{fig:cr_vs_mhalo_halomassbins} shows X-ray emission in excess of random stacking expectations with SNR$>$4 in {the three bins spanning $M_{\rm halo} = 10^{12.6-14.5}\, M_\odot$.} 

{After masking AGN, the detections in these bins are maintained with lower SNR, although the $10^{13.3-14}\, M_\odot$ bin detection is still marginal as above. As in our Figure \ref{fig:hardband_all_stack} analysis, comparing AGN-masked count rates and XRB expectations in Figure \ref{fig:cr_vs_mhalo_halomassbins} shows that (i) only the $M_{\rm halo} = 10^{14-14.5}\, M_\odot$  bin has an AGN-masked count rate that exceeds all 13 XRB estimators we considered, with our preferred XRB estimator accounting for just $6\pm1\%$ of the observed AGN-masked count rate, and (ii) our preferred XRB estimator accounts for just $24\pm5\%$ and $30\pm4\%$, respectively, of the AGN-masked count rates in the two lower $M_{\rm halo} = 10^{13.3-14}\, M_\odot$ and $M_{\rm halo} = 10^{12.6-13.3}\, M_\odot$ bins, indicating likely hot gas detections in these bins as well. Although other XRB estimators among the 13 we considered could fully explain these bins’ AGN-masked emission, the same estimators would overpredict the emission in lower mass bins.}

The right panel of Figure \ref{fig:cr_vs_mhalo_halomassbins} shows the corresponding distribution of $M_{\rm HI,grp}/M_{\rm halo}$ as a function of halo mass. Group \HI{} content smoothly declines by 1.2 dex over the same halo mass range in which X-ray counts increase.  {Based on expected XRB contributions, our detections of diffuse X-rays from hot gas become confident across $M_{\rm halo} = 10^{12.6-14}\, M_\odot$ and unambiguous above $M_{\rm halo} = 10^{14}\, M_\odot$}, where $M_{\rm HI,grp}/M_{\rm halo}$ simultaneously reaches its lowest values as a function of halo mass. Together, these results suggest an inverse relationship between halo-integrated \HI{} content and halo-integrated X-ray emission, reflecting both hot gas and galactic X-ray sources. This relationship is discussed further in \S\ref{sec:discussion}.

\begin{figure*}
    \centering
    \includegraphics[scale=0.9]{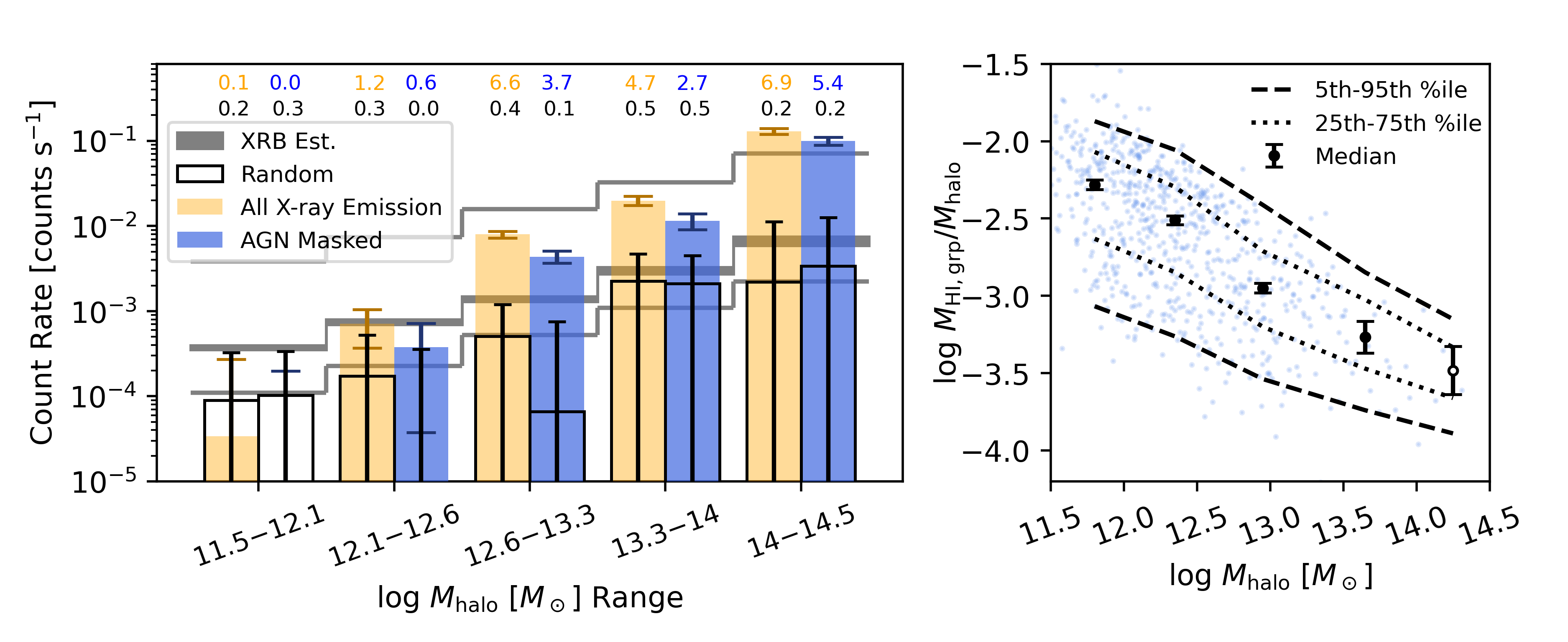}
    \caption{Group stacked X-ray emission and $M_{\rm HI,grp}/M_{\rm halo}$ in bins of group halo mass, as presented in the bottom row of Figure \ref{fig:hardband_all_stack}, but now excluding groups with $N_{\rm galaxies}=1$ and/or $M_{\rm halo} < 10^{11.5}\, M_\odot$. The annotated numbers in the right panel give $n_{\rm bin}$ values for each bin, which are the same for both left and right panels. In the right panel, errors on the data points come from bootstrapping, and open circles represent bins with fewer than 30 groups, for which bootstrap errors may be unreliable.}
    \label{fig:cr_vs_mhalo_halomassbins}
\end{figure*}


\subsection{How do group cold gas and X-ray emission depend on virialization state at fixed halo mass?}\label{subsec:tcrossresults}
With their halo mass dependence established, we can now assess how group \HI{} content and X-ray emission depend on virialization state, as parameterized by crossing time, at fixed halo mass. Figure \ref{fig:tcrossmotivation} shows groups distributed in $M_{\rm HI,grp}/M_{\rm halo}$ versus $t_{\rm cross}$, in three panels representing bins of group halo mass. Analysis of $t_{\rm cross}$ requires $N_{\rm galaxies}>1$ groups, so we again exclude $M_{\rm halo} < 10^{11.5}\, M_\odot$. For $M_{\rm halo}=10^{11.5-12.1}\,M_\odot$ and $M_{\rm halo}=10^{12.1-12.6}\,M_\odot$, Figure \ref{fig:tcrossmotivation} shows a transition to greater spread in $M_{\rm HI,grp}/M_{\rm halo}$ as crossing time decreases, with many more low-$M_{\rm HI,grp}/M_{\rm halo}$ values appearing despite the upper end of the range not changing. We see this transition occurring across $\log\left[t_{\rm cross}/13.8\,{\rm Gyr}\right]\sim -0.86$ (marked with vertical line), which is similar to the median values of $\log\left[t_{\rm cross}/13.8\,{\rm Gyr}\right]$ in the bins shown ($-0.89$, $-0.85$, $-0.86$, respectively) and represents $\sim$2 Gyr. The existence of this transition motivates us to ask whether X-ray emission increases below the same $t_{\rm cross}$ value, possibly due to virial shock heating and/or environmental triggering of star formation or AGN.

\begin{figure*}
    \centering
    \includegraphics[scale=0.87]{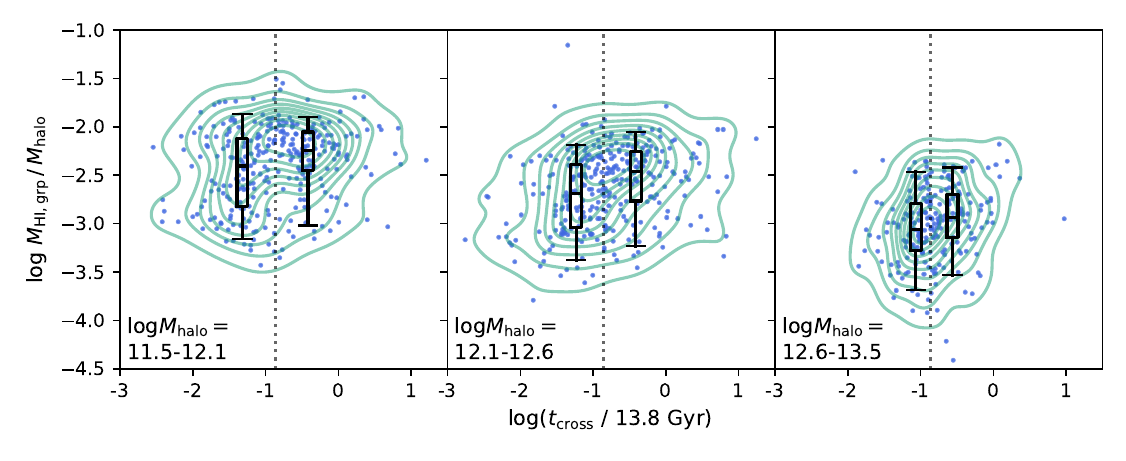}
    \caption{Group-integrated \HI{}-to-halo mass ratios ($M_{\rm HI,grp}/M_{\rm halo}$) as a function of group crossing time, expressed as a fraction of the age of the Universe, in three halo mass bins. Contour lines have been drawn using kernel density estimation. The vertical line corresponds to $\log(t_{\rm cross} / {\rm 13.8\, Gyr}) = -0.86$ ($\sim$2 Gyr), as used in \S\ref{subsec:tcrossresults} to categorize groups as lower-$t_{\rm cross}$ or higher-$t_{\rm cross}$. {Box plots show the distributions of $M_{\rm HI,grp}/M_{\rm halo}$ above and below $\log(t_{\rm cross} / {\rm 13.8\, Gyr}) = -0.86$ in each panel. The solid bar represents the median value, the box represents the interquartile range, and the whiskers show the 5$^{\rm th}$--95$^{\rm th}$ percentiles.} At crossing times smaller than $\log(t_{\rm cross} / {\rm 13.8\, Gyr}) = -0.86$, groups show greater diversity in \HI{} content (particularly at lower $M_{\rm halo}$), as discussed in \S\ref{subsec:tcrossresults}.}
    \label{fig:tcrossmotivation}
\end{figure*}

To address this question, we provide two figures connecting group \HI{} content (Figure \ref{fig:hi_tcross_mhalo}) and X-ray emission (Figure \ref{fig:tcross_stacking}) to $t_{\rm cross}$. In Figure \ref{fig:hi_tcross_mhalo}, we show trends of median $M_{\rm HI,grp}/M_{\rm halo}$ versus $M_{\rm halo}$ for lower-$t_{\rm cross}$ and higher-$t_{\rm cross}$ groups, which are separated at $\log(t_{\rm cross} / {\rm 13.8\, Gyr}) = -0.86$ based on Figure \ref{fig:tcrossmotivation}.\footnote{For both Figures \ref{fig:hi_tcross_mhalo} and \ref{fig:tcross_stacking}, our results do not noticeably change when adopting boundary values of $-0.75$ or $-1$ instead of our chosen value of $-0.86$.} Lower-$t_{\rm cross}$ groups exhibit lower median $M_{\rm HI,grp}/M_{\rm halo}$ than higher-$t_{\rm cross}$ groups at fixed halo mass, with a maximum difference of 0.2 dex (a factor of $\sim$1.6). We note that while Figure \ref{fig:hi_tcross_mhalo} includes X-ray confused groups in calculations (whereas Figure \ref{fig:tcross_stacking} excludes these groups), the trends in Figure \ref{fig:hi_tcross_mhalo} are not noticeably different if these confused groups are excluded.

\begin{figure}
    \centering\includegraphics{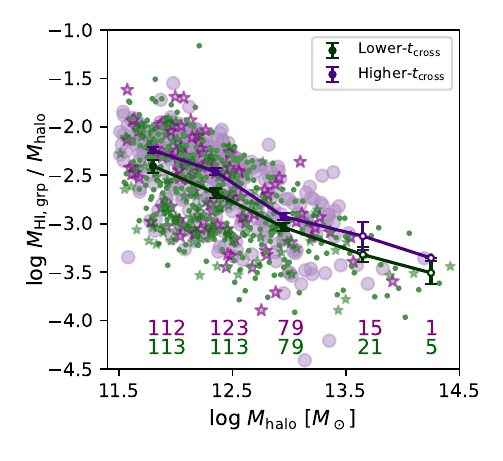}
    \caption{Median $M_{\rm HI,grp}/M_{\rm halo}$ vs. $M_{\rm halo}$ for lower-$t_{\rm cross}$ (green) and higher-$t_{\rm cross}$ (purple) groups, as defined in \S\ref{subsec:tcrossresults}.  We exclude $N_{\rm galaxies}=1$ groups from this plot.  Error bars on the medians were computed using bootstrapping with 5,000 resamples; {open circles represent bins that contain too few ($<$30) groups to attain reliable bootstrapped errors.} X-ray confused groups are marked with stars rather than points. The numbers of groups within each bin, excluding X-ray confused groups (to match Figure \ref{fig:tcross_stacking}), are annotated at the bottom. }
    \label{fig:hi_tcross_mhalo}
\end{figure}

Figure \ref{fig:tcross_stacking} illustrates stacked X-ray count rates for galaxy pairs and groups in the same $M_{\rm halo}$ bins and $t_{\rm cross}$ categories as in Figure \ref{fig:hi_tcross_mhalo}.  With AGN not masked, as shown in the left panel, we find SNR$>$4.5 detections for both $t_{\rm cross}$ categories at $M_{\rm halo} = 10^{12.6-13.3}\, M_\odot$. In the $M_{\rm halo} = 10^{13.3-14}\, M_\odot$  and $M_{\rm halo} = 10^{14-14.5}\, M_\odot$ bins, we find weak SNR$\sim$2.8 detections for higher-$t_{\rm cross}$ bins and significant SNR$\geq$4.9 detections for lower-$t_{\rm cross}$ bins. The measured count rate exceeds random stacking expectations in all cases. {We note that if we exclude group \#1345 (see Footnote \ref{footnote:ECO04631})  from the higher-$t_{\rm cross}$ $M_{\rm halo}=10^{12.6-13.3}\,M_\odot$ bin, the SNR of that stack drops to 4.}

The SNRs of these detections drop when AGN are masked, {as illustrated in the right panel}, and in this case we still find {SNR$\geq$2.9} detections in excess of random stacks for higher-$t_{\rm cross}$ groups at $M_{\rm halo} = 10^{12.6-13.3}\, M_\odot$ and for lower-$t_{\rm cross}$ groups in the two bins spanning $M_{\rm halo} = 10^{13.3-14.5}$. There is additionally a weaker {SNR$\sim$2.2} detection for lower-$t_{\rm cross}$ groups at $M_{\rm halo} = 10^{12.6-13.3}\, M_\odot$, consistent with XRB expectations. The most significant detection is for lower-$t_{\rm cross}$ groups at $M_{\rm halo} = 10^{14-14.5}\, M_\odot$, where the SNR reaches {5.9}.

With these results, we cannot confidently ascertain whether a general relationship exists between group X-ray emission and group crossing time at fixed halo mass.  There is a hint, however, that enhanced group X-ray emission from hot gas or galactic sources is associated with greater virialization (lower-$t_{\rm cross}$) above $M_{\rm halo} = 10^{13.3}\, M_\odot$. With AGN masked, the low-$t_{\rm cross}$ bins at $M_{\rm halo} = 10^{13.3-14}\, M_\odot$ and $M_{\rm halo} = 10^{14-14.5}\, M_\odot$ show enhanced count rates, of which {(based on our preferred XRB estimator) XRBs are expected to contribute $22\pm4\%$ and $4\pm1\%$, respectively.} This result flips in the lower $M_{\rm halo} = 10^{12.6-13.3}\, M_\odot$ halo mass bin, where higher-$t_{\rm cross}$ groups show enhanced count rate. In this case, XRBs contribute {$15\pm3\%$}. Deeper X-ray data over a larger number of groups would be needed to confirm and better understand these results. The binning of our sample by both $M_{\rm halo}$ and $t_{\rm cross}$, on top of excluding $N_{\rm galaxies}=1$ halos and confused halos, leaves us $<$125 groups per stacking bin, thereby reducing our stacking depth and increasing our sensitivity to outliers. We further discuss these results and their connection to group \HI{} content in \S\ref{subsec:discuss:vir}.

\begin{figure*}
    \centering
    \includegraphics[scale=0.22]{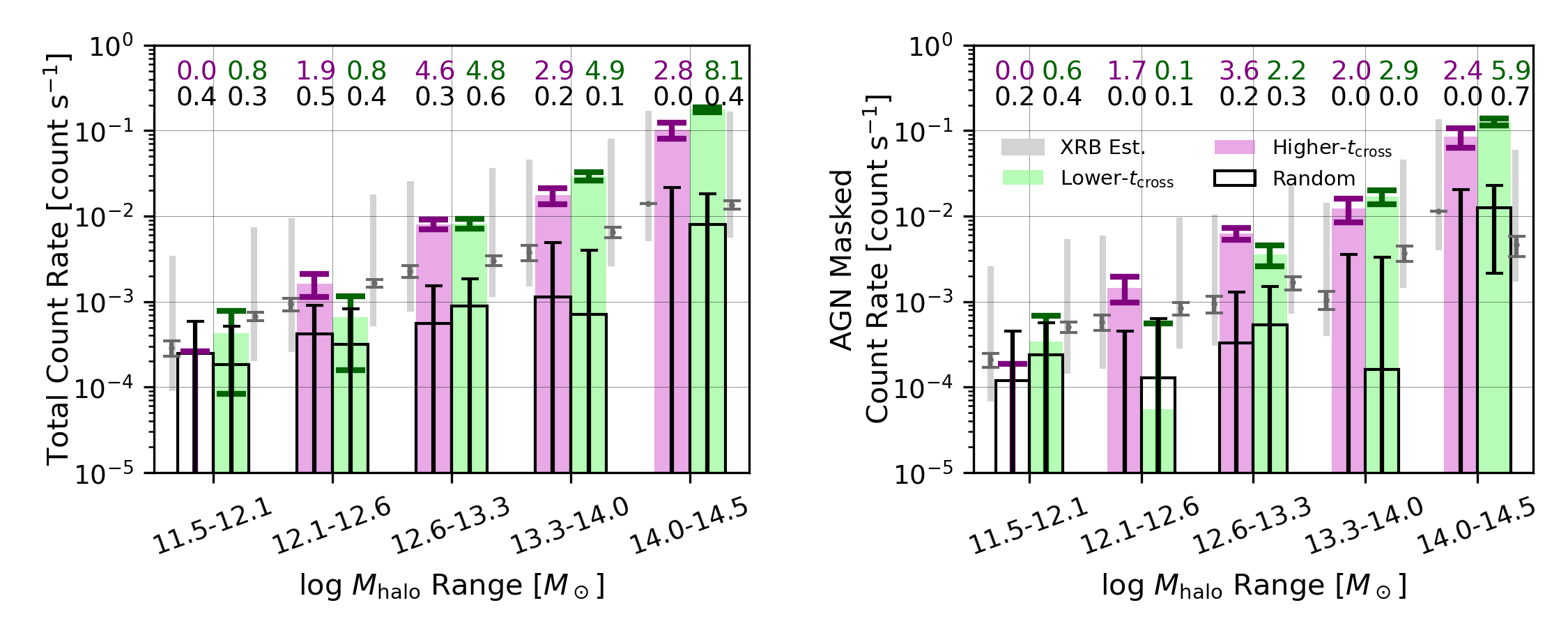}
    \caption{Stacked X-ray emission in bins of group halo mass separated into higher- and lower-$t_{\rm cross}$ categories (see \S\ref{subsec:tcrossresults}) {Left:} Total count rate as a function of halo mass for lower-$t_{\rm cross}$ (green) and higher-$t_{\rm cross}$ (pink) groups. Black lines represent count rates for random stacks. {Gray points and error bars represent our preferred XRB estimate and its uncertainty from bootstrapping, while the gray bars represent the range between the minimum and maximum XRB estimator for each stack.} Numbers annotated at the top are SNRs for each stacking bin, with SNRs corresponding to random stacking in black. {Right: } Same as left panel, but count rates are measured with AGN masked.}
    \label{fig:tcross_stacking}
\end{figure*}

\subsection{How do group cold gas content and X-ray emission depend on AGN and SF at fixed halo mass?}\label{subsec:agnresults}

Figures \ref{fig:agn_hi_sfr}--\ref{fig:agn_tcross_connection} assess how AGN and SF relate to cold gas content and X-ray emission. Figure \ref{fig:agn_hi_sfr} shows trends of median $M_{\rm HI,grp}/M_{\rm halo}$ and $\rm FSMGR_{\rm grp}$, dividing our sample into AGN-hosting and non-AGN-hosting halos. $N_{\rm galaxies}=1$ halos are included. Trendlines are computed in sliding 0.3-dex halo mass bins.   {The sliding statistic helps us better identify fine structure in the data that might be suppressed by discrete binning.} However, we also superpose points with error bars (derived from bootstrapping) for independent 0.3-dex bins. 

\begin{figure}
    \centering
    \includegraphics[scale=0.9]{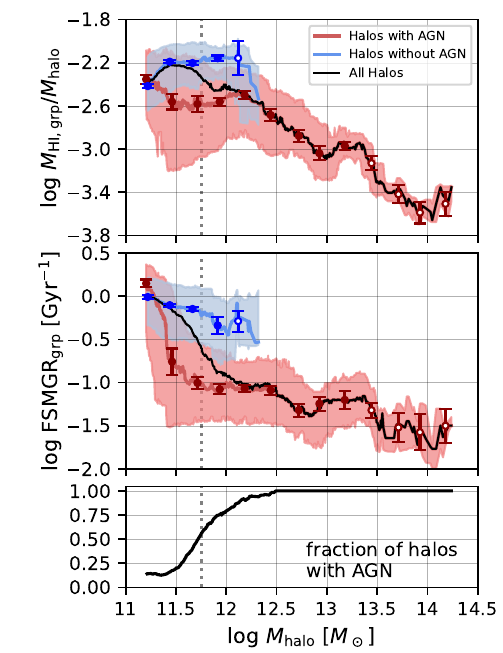}
    \caption{Median $M_{\rm HI,grp}/M_{\rm halo}$, median \fsmgrgrp, and fraction of halos with AGN as a function of halo mass. All panels include $N_{\rm galaxies} = 1$ groups. {Top:} Median $M_{\rm HI,grp}/M_{\rm halo}$ vs. $M_{\rm halo}$ for halos with AGN (red) and without AGN (blue). Lines were generated using a sliding window, {which advances forward in one-data-point increments} to compute the median $M_{\rm HI,grp}/M_{\rm halo}$ and median $M_{\rm halo}$ values in {overlapping} 0.3-dex bins of halo mass. {Data points show medians in independent 0.3-dex bins extracted from these sliding median trendlines, with error bars determined by bootstrapping; open data points denote bins for which the bootstrapped error is unreliable due to having $<$30 data points in the bin.} As a consequence of the different sampling of AGN-hosting and non-AGN-hosting halos as a function of halo mass, the data points for the AGN-hosting and non-AGN-hosting selections do not fall at exactly the same halo mass values. Shaded regions represent the interquartile (middle 50\%) range, also computed in a sliding window. {Middle:} Median group-integrated FSMGR vs. $M_{\rm halo}$ for halos with and without AGN. Lines, colors, points, and shaded regions are as in the top panel. We note that the elevation of the lowest-mass \fsmgrgrp{} point for AGN-hosting halos is a binning artifact and not a robust result. {Bottom:} Fraction of AGN-containing halos as a function of $M_{\rm halo}$, computed using a sliding 0.3 dex window in group halo mass.}
    \label{fig:agn_hi_sfr}
\end{figure}

Figure \ref{fig:agn_hi_sfr} demonstrates that below the bimodality scale of $M_{\rm halo} = 10^{12.1}\, M_\odot$, halos with AGN have reduced median $M_{\rm HI,grp}/M_{\rm halo}$ {and \fsmgrgrp{}} compared to halos without AGN.\footnote{{Given that our calculated FSMGRs (see \S\ref{subsubsec:data:g3groups}) could reflect potential systematics associated with SED fitting \citep{lower2020well}, we have confirmed that our result still holds if we instead calculate \fsmgrgrp{} using UV and mid-IR photometry from M.S. Polimera et al. (2025, in prep.). These alternative FSMGRs have been calculated using standard composite SFR prescriptions \citep{buat2011spectral}, which typically assume a constant star formation history over a relatively short timescale $\sim$100 Myr, and they rely on SED fitting only in the use of internal extinction corrections.}} The reduced $M_{\rm HI,grp}/M_{\rm halo}$ for AGN-hosting halos is significant to $>$5$\sigma$ according to a two-sample Kolmogorov-Smirnov test, and it reaches a depth of $\sim$0.4 dex at $\log M_{\rm halo} = 11.5 \pm 0.1$. {Within this halo mass range, non-AGN-hosting halos exhibit a plateau in $M_{\rm HI,grp}/M_{\rm halo}$, whereas} AGN-hosting halos exhibit a broad valley in $M_{\rm HI,grp}/M_{\rm halo}$ over {1 dex} in halo mass centered on {$\log M_{\rm halo} = 11.8 \pm 0.1$} (based on a parabolic fit to the valley {labeled ``a'' in Figure \ref{fig:hihmdip}}). {Valley ``a''}  underlies the noticeable yet narrower depression in $M_{\rm HI,grp}/M_{\rm halo}$ ratios {labeled ``b'' that is} seen for all halos (black line), which occurs across $M_{\rm halo} \sim 10^{11.6-12}\, M_\odot$ as the fraction of halos with AGN rises, crossing 50\% at {$\log M_{\rm halo} \sim 10^{11.7}\, M_\odot$}. To assess the statistical significance of valley ``b,'' we constructed a baseline for our median $M_{\rm HI,grp}/M_{\rm halo}$ vs. $M_{\rm halo}$ data by fitting to the analytic model of \citet{obuljen2019hi}, resulting in $\log M_0 = 9.4$, $\log M_{\rm min} = 11.3$, and $\alpha = 0.39$ (see also H23), then divided the integrated area above the valley into the integrated noise expected from our uncertainties in the median values. This calculation indicates a confidence of 2.5$\sigma$ for valley ``b'' across $M_{\rm halo} = 10^{11.6-12.1}\, M_\odot$. While we lack a model baseline to quantify the larger valley ``a'' seen for AGN-hosting halos taken alone, the fact that it represents a $>$5$\sigma$ deviation from non-AGN-hosting halos provides a physical explanation for valley ``b'' and compels us to consider both valleys meaningful.

{Above the bimodality scale, we note another potential valley in the trend for AGN-hosting halos located at approximately $M_{\rm halo} \sim 10^{12.7 - 13.3} \, M_\odot$ (labeled ``c'' in Figure 11), which is mirrored in the trend for all halos. This high-mass valley deviates from our model fit to the median $M_{\rm HI,grp}/M_{\rm halo}$ vs. $M_{\rm halo}$ relation (see above) with a significance of 3.9$\sigma$; however, this deviation could also be interpreted as a 1.3$\sigma$ bump at $\log M_{\rm halo} \sim 13.3$ rather than a  valley at $\log M_{\rm halo} \sim 13.0$. If the high-mass valley labeled ``c'' in Figure 11 is truly a valley rather than a positive bump, it has higher significance than the narrow valley labeled ``b,''  which has a demonstrable physical origin in the (larger and more significant valley) labeled ``a.''} {Finally, given concerns about the reliability of mid-IR AGN classifications in the dwarf regime (e.g., \citealp{hainline2016mid, sturm2025star}), we note that our results in Figure \ref{fig:agn_hi_sfr} do not noticeably change if we treat mid-IR AGN host galaxies as non-AGN-hosting galaxies, due to their relative infrequency (2.5\% of our AGN).}

\begin{figure*}
    \centering
    \includegraphics[scale=0.87]{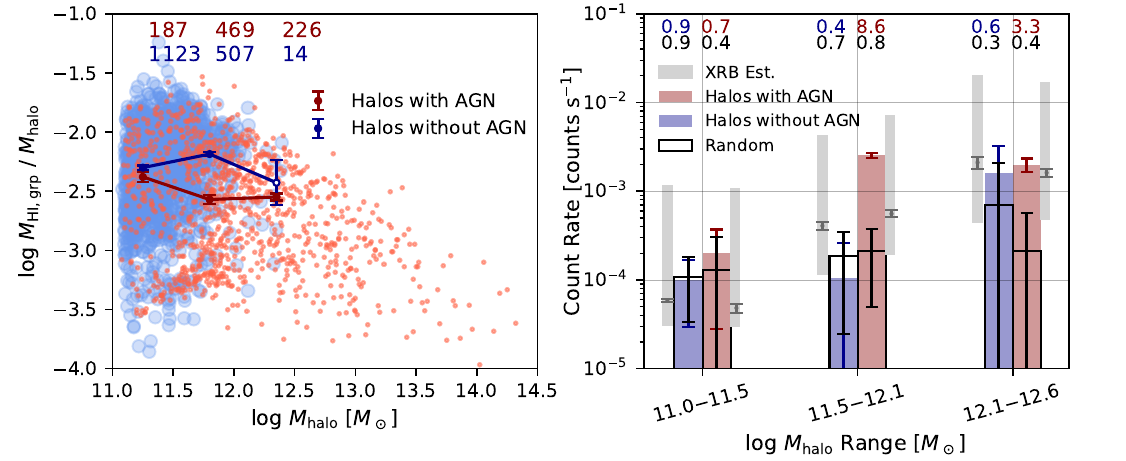}
    \caption{$M_{\rm HI,grp}/M_{\rm halo}$ and stacked X-ray emission for AGN-hosting and non-AGN-hosting halos in fixed halo mass bins. {Left: } Median $M_{\rm HI,grp}/M_{\rm halo}$ vs. $M_{\rm halo}$ for AGN-hosting halos (red) and non-AGN-hosting halos (blue), following Figure \ref{fig:agn_hi_sfr} but now with halo mass bins matched to our X-ray stacks in the right panel. Open points denote bins for which the bootstrapped error bars are unreliable due to having $<$30 points. $N_{\rm galaxies}=1$ halos are included. The numbers annotated at the top are the numbers of halos within each stacking bin, as used in the right panel. X-ray confused groups are excluded for consistency with the right panel. {Right: } Stacked background-subtracted X-ray count rate vs. $M_{\rm halo}$ for AGN-hosting and non-AGN-hosting halos. Count rates are measured with RESOLVE and ECO AGN not masked. Annotated numbers are SNRs for the count rate measured in each stacking bin. {Black points show our preferred XRB estimate based on \citet{lehmer2019x}, and gray bars show the range of XRB estimates across all estimators (see \S\ref{subsec:xrbmethod}).}}
    \label{fig:agnstackingplot}
\end{figure*}

To investigate whether these relationships may have associated X-ray emission signatures reflecting galactic X-ray sources or hot gas, Figure \ref{fig:agnstackingplot} shows background-subtracted X-ray count rates  versus halo mass for AGN-hosting and non-AGN-hosting halos, measured with AGN unmasked. We have stacked halos in bins up to {$M_{\rm halo} = 10^{12.1-12.6}\, M_\odot$, the highest mass bin that contains non-AGN-hosting halos. For non-AGN-hosting halos, we do not detect significant X-ray emission in any of the bins. However, for AGN-hosting halos, we detect SNR$>$3 X-ray emission in excess of random stacking expectations in the two halo mass bins spanning $M_{\rm halo} = 10^{11.5-12.6}\, M_\odot$ but not from the lower mass bin at $M_{\rm halo}=10^{11-11.5}\, M_\odot$. To test whether the excess emission arises from or near the AGN, we replicated the analysis of Figure \ref{fig:agnstackingplot} but instead masked known AGN (not pictured). The detections in the $M_{\rm halo} = 10^{11.5-12.1}\, M_\odot$ and $M_{\rm halo} = 10^{12.1-12.6}\, M_\odot$ bins dropped to SNR$<$2. This result confirms that some of the excess X-rays we measure in Figure \ref{fig:agnstackingplot} derive from AGN directly or from hot gas (possibly heated by AGN).}


\subsection{Connecting crossing time and AGN results}\label{subsec:results:connecting}
Finally, we consider whether our separate results connecting group \HI{} content to crossing time and to AGN/SF are related, as might occur if group assembly lowers $t_{\rm cross}$ and thereby increases the likelihood of triggering AGN or enhancing SF. This analysis requires that we exclude $N_{\rm galaxies}=1$ groups to calculate $t_{\rm cross}$, leaving only 410 AGN-hosting groups and only 30 non-AGN-hosting groups at all halo masses. Given this small number of non-AGN-hosting groups as well as our inconclusive \S\ref{subsec:tcrossresults} X-ray stacking results when binning in $M_{\rm halo}$ and $t_{\rm cross}$, we do not consider X-ray emission in this joint analysis. 

Figure \ref{fig:agn_tcross_connection} shows the $M_{\rm HI,grp}/M_{\rm halo}$ vs. $M_{\rm halo}$ relation subdivided into four bins, representing AGN-hosting and non-AGN-hosting halos as well as lower-$t_{\rm cross}$ and higher-$t_{\rm cross}$ groups. Lines represent median values in sliding windows and point represent medians and errors in independent {0.4} dex bins on the sliding median trendline; as in Figure \ref{fig:agn_hi_sfr}, the points may not overlap since for different subsamples the sliding medians start and end at different halo masses.

The left panel illustrates that AGN-hosting halos at higher $t_{\rm cross}$ have elevated $M_{\rm HI,grp}/M_{\rm halo}$ compared to AGN-hosting halos at lower $t_{\rm cross}$. At $M_{\rm halo}\sim 10^{12.5-13}\, M_\odot$ and higher masses, the median trendlines for higher- and lower-$t_{\rm cross}$ AGN-hosting halos become consistent, given the error bars on the points. {In the right panel, we see that low-$t_{\rm cross}$ and high-$t_{\rm cross}$ non-AGN-hosting halos have similar $M_{\rm HI,grp}/M_{\rm halo}$, albeit with the caveat that the bootstrapped uncertainties on $M_{\rm HI,grp}/M_{\rm halo}$ may not be reliable given the small number of $N_{\rm galaxies}>1$ groups that lack AGN. Thus, a larger sample of non-AGN-hosting halos would be helpful for ascertaining whether the relationships connecting group \HI{} content to group crossing time and AGN presence are independent. }

\begin{figure*}
    \centering
    \includegraphics[scale=0.9]{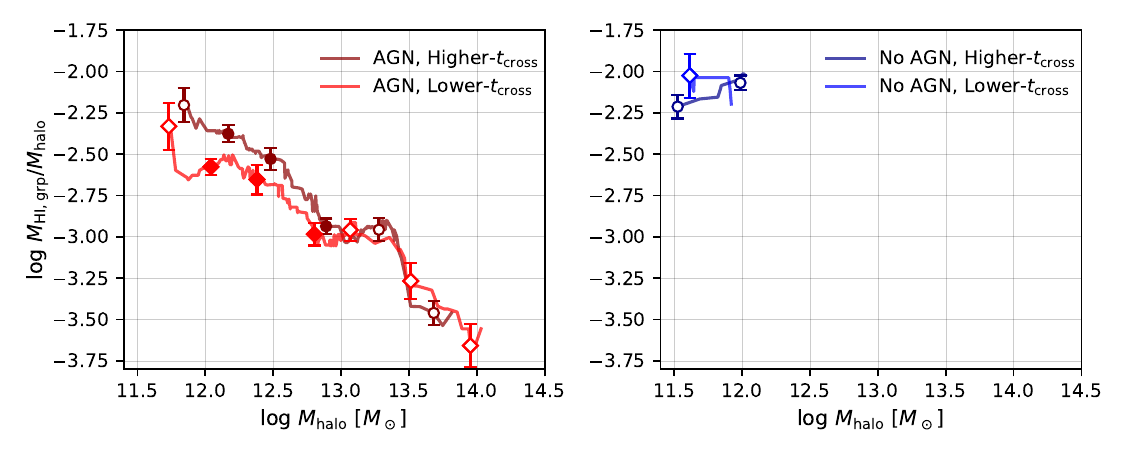}
    \caption{Demonstration of the independent effects of AGN and $t_{\rm cross}$ on $M_{\rm HI,grp}/M_{\rm halo}$. {Left: } Median $M_{\rm HI,grp}/M_{\rm halo}$ vs. $M_{\rm halo}$ for higher-$t_{\rm cross}$ (thin line) and lower-$t_{\rm cross}$ (thick line) AGN-hosting halos. Lines, points, and error bars are as in Figure \ref{fig:agn_hi_sfr}. $N_{\rm galaxies}=1$ halos are excluded. {Right: } Same as left panel, but for non-AGN-hosting halos.
    }
    \label{fig:agn_tcross_connection}
\end{figure*}

\section{Discussion}\label{sec:discussion}
Our results in the previous section connect the relationship between group \HI{} content and X-ray emission to group crossing time, star formation, and AGN activity. We now discuss broader implications and compare to past observational and theoretical work. We first discuss the gas inventory in halos, then discuss possible physical scenarios relating virialization state, AGN, and star formation to halo \HI{} content.

\subsection{The Gas Inventory in Halos}\label{subsec:discuss:gasinventory}
In \S\ref{subsec:logmhstacking}, our results implied an inverse relationship between group \HI{} content and the presence of hot intragroup gas. {At high halo masses $M_{\rm halo} \gtrsim 10^{12.6}\, M_\odot$, groups have low \HI{}-to-halo mass ratios, and their stacked X-ray count rates are unlikely to be fully explained by X-ray binaries and AGN,\footnote{{Dwarf AGN appear to be typically underluminous in X-rays \citep[M.S. Polimera et al. 2025, in prep.]{latimer2019x}. Moreover, among dwarf AGN detected using all the aforementioned methods, the ``SF-AGN'' identified by \citet{polimera2022resolve} are the most abundant and have the most  dwarf-like host galaxy properties, i.e., metal-poor, gas-rich, and star-forming. The abundance of these ``typical dwarf'' AGN exceeds that of X-ray-detected dwarf AGN by at least a factor of five (comparing \citealp{polimera2022resolve} to \citealp{birchall2022incidence}), so we do not expect an additional undetected dwarf AGN population with significant X-ray emission to contribute to our stacks, although we cannot rule it out.}} suggesting the presence of diffuse hot gas. {We found that lower-mass halos have higher \HI{}-to-halo mass ratios and stacked X-ray emission that can probably be explained in full by AGN or XRBs.

X-ray detections at these high halo masses are expected from previous work. In a stacking analysis using RASS data, \citet{anderson2012extended} detected strong X-ray emission from both early- and late-type luminous, isolated galaxies but reported that faint, isolated galaxies (central galaxy $K_s \geq -24.1$, typically $M_{\rm halo}\sim 10^{12.3}\, M_\odot$ in ECO) showed no evidence of extended emission. In a larger {RASS stacking analysis of} $\sim$250,000 SDSS brightest cluster galaxies, \citet{anderson2015unifying} further detected extended hot gas down to central galaxy $M_* = 10^{10.8}\, M_\odot$, corresponding to $M_{\rm halo} = 10^{12.6}\, M_\odot$. Furthermore, deeper X-ray observations using \textsl{Chandra}, \textsl{XMM-Newton}, and \textsl{eROSITA} have measured hot gas luminosity vs. group halo mass down to $M_{\rm halo} \sim 10^{11}\, M_\odot$ \citep{kim2013scaling,goulding2016massive,forbes2017sluggs,bogdan2022x,zhang2024hot}. The steep slope of this power law highlights the much weaker X-ray emission of lower-mass halos in comparison to higher-mass halos. {As such, our stacking analysis, despite being based on volume-limited, dwarf-dominated surveys, may have lacked enough low-mass halos (compared to these other studies) to detect hot gas at these low halo masses.}

Scarcity of hot gas in low-mass halos would be consistent with both theoretical and observational expectations that most halo gas in these halos is in the form of the warm-hot {intergalactic} medium (WHIM). In simulations, WHIM constitutes 40--50\% of the baryonic mass inventory at temperatures and densities that emit only weakly in X-rays \citep{cen1999baryons, dave2001baryons,smith2011nature}. Examining the cold baryonic mass (stars + atomic gas) function in RESOLVE and ECO and combining it with literature prescriptions for hot halo gas and galactic molecular gas, \citet{eckert2017baryonic} found a mass deficit below $M_{\rm halo}=10^{12.1}\,M_\odot$ between the observed baryonic mass function and the expected baryonic mass function, based on the halo mass function and assuming a uniform baryonic fraction. This deficit matched theoretical estimates for WHIM in low-mass halos.

\subsection{Group Virialization State}\label{subsec:discuss:vir}

In \S\ref{subsec:tcrossresults}, we demonstrated that lower-$t_{\rm cross}$ galaxy groups show statistically lower $M_{\rm HI,grp}/M_{\rm halo}$ ratios and greater spread in $M_{\rm HI,grp}/M_{\rm halo}$ than higher-$t_{\rm cross}$ groups at fixed halo mass. The transition to greater spread in group \HI{} content occurs across $\log(t_{\rm cross} / 13.8\,{\rm Gyr}) = -0.86$, or $\sim$2 Gyr. It is interesting that this transition becomes prominent for $M_{\rm halo} = 10^{12.1-12.6}\, M_\odot$ and even more so for $M_{\rm halo} = 10^{11.5-12.1}\, M_\odot$, whereas at higher halo masses there is a smoother relationship between $M_{\rm HI,grp}/M_{\rm halo}$ and $t_{\rm cross}$ across this $\sim$2 Gyr scale (see Figure \ref{fig:tcrossmotivation}).

Using a sample of 172 SDSS groups, \citet{ai2018evolution} have also examined the relationship between $M_{\rm HI,grp}/M_{\rm halo}$ and $t_{\rm cross}$. Their results do not appear to show such a transition to greater spread in $M_{\rm HI,grp}/M_{\rm halo}$ across $\log(t_{\rm cross} / 13.8\,{\rm Gyr}) = -0.86$, perhaps due to the higher halo masses of their groups, which span $M_{\rm halo} \sim 10^{13-14.5}\, M_\odot$. However, our results are consistent with theirs in showing suppressed median $M_{\rm HI,grp}/M_{\rm halo}$ for groups with lower $t_{\rm cross}$. Through comparison of crossing time to \HI{} depletion timescales, the authors argued that this relationship implies that long-timescale processes (e.g., starvation; \citealp{bekki2002passive}) are more important in group evolution than short-timescale environmental quenching processes (e.g., ram-pressure stripping; \citealp{abadi1999ram}). These processes are usually associated with massive groups and clusters (due to their being more effective in environments with hot gas), but their effects have been seen in less massive halos around the bimodality scale and even below it (e.g., \citealp{grcevich2009h, li2017farthest,putman2021gas,zhu2023census,jones2024gas}). Thus the continued trend we see down to $M_{\rm halo}=10^{11.5-12}\,M_\odot$ (Figure \ref{fig:hi_tcross_mhalo}) is not surprising.

In addition, mergers are expected to be common in halos typical of small groups, given the more effective dynamical friction at low peculiar velocities \citep{chandrasekhar1943dynamical}. Thus, the scatter toward low $M_{\rm HI,grp}/M_{\rm halo}$ in halos with low $t_{\rm cross}$ could alternatively reflect increased rates of \HI{} processing in more compact halos. As groups collapse, mergers and interactions may become more frequent or more intense, and prior observations indicate these interactions may trigger atomic gas consumption and associated feedback that may also deplete gas. {Many studies have directly linked interactions and minor mergers to enhancement of AGN activity and SF (e.g., \citealp{di2008frequency, stark2013fueling, pipino2014zurich, kaviraj2014importance, comerford2015merger, gao2020mergers})}. {It is possible, however, that the relationships connecting group \HI{} to $t_{\rm cross}$ and AGN presence are independent; our result in Figure \ref{fig:agn_tcross_connection} includes too few non-AGN-hosting groups for us to say conclusively.}

AGN aside, our X-ray emission results complicate the story. If lower $t_{\rm cross}$ truly indicates virialization state, AGN-masked X-ray count rates are greater for more-virialized groups at $M_{\rm halo} > 10^{13.3}\, M_\odot$ but greater for less-virialized groups at $M_{\rm halo} = 10^{12.6-13.3}\, M_\odot$. These results follow expectations at high halo mass: above the ``shutdown'' scale at $\sim10^{13.3}\, M_\odot$, models predict full halo gas heating to multiple virial radii \citep{dekel2006galaxy} and observations find that nearly all galaxies are gas-poor and quenched \citep{kannappan2009s0, kannappan2013connecting, moffett2015eco}. 

The opposite result for $M_{\rm halo} = 10^{12.6-13.3}\, M_\odot$ is an open puzzle. We note that crossing time is an imperfect metric of virialization state, and our broad categories of higher-$t_{\rm cross}$ versus lower-$t_{\rm cross}$ may oversimplify the full diversity of group \HI{} content as a function of halo mass and virialization state. Groups with $M_{\rm halo} = 10^{12.6-13.3}\, M_\odot$ are relatively small, so discrete merger events may make $t_{\rm cross}$ noisy. As group galaxies merge and the group evolves towards becoming a ``fossil group'' (e.g., \citealp{ponman1994possible}), the group \HI{} content will drop (see \citealp{guo2020direct}) but the group crossing time may not smoothly decrease. In fact, since galaxy mergers can remove some of the smallest group-relative on-sky distances from the $t_{\rm cross}$ calculation, crossing time could \emph{increase} (contrary to expectations for increased virialization) after a merger. This scenario exemplifies that crossing time may not perfectly capture how \HI{} content relates to group assembly, especially for small groups in which low velocity dispersions enable enhanced merging (e.g., \citealp{carlberg2001environment}). Given this caveat and the additional complication that $t_{\rm cross}$ is subject to projection effects, future work on this topic may benefit from incorporating additional virialization metrics sensitive to merging history (e.g., magnitude gap; \citealp{trevisan2017finer}).

\subsection{AGN and Star Formation}\label{subsub:discuss:agn}
One of our key results in \S\ref{subsec:agnresults} (Figure \ref{fig:agn_hi_sfr}) is the connection of AGN to halo-integrated \HI{} content and FSMGR. At low halo masses $M_{\rm halo} \lesssim 10^{12.1}\, M_\odot$, AGN-hosting halos show a broad valley in median $M_{\rm HI,grp}/M_{\rm halo}$ and median $\rm FSMGR_{\rm grp}$ compared to non-AGN-hosting halos at fixed halo mass. {The trend that non-AGN-hosting halos have higher median $M_{\rm HI,grp}/M_{\rm halo}$ and \fsmgrgrp{} continues just past the bimodality scale to $M_{\rm halo}\sim 10^{12.3}\, M_\odot$, above which all of our halos host AGN. At higher masses, we find a possible valley at $M_{\rm halo}\sim 10^{13}\, M_\odot$, though whether it has a relationship to AGN is unknown. Interestingly,} \citet{ellision2019atomic} found that the \HI{}-richness of AGN host galaxies is statistically lower than that of non-AGN-hosts below galaxy $M_*\sim 10^{10.25}\, M_\odot$ (corresponding to $M_{\rm halo} = 10^{11.9} M_\odot$ centrals in ECO), with opposite behavior at higher stellar masses}. {In this section, we discuss possible scenarios that may account for these trends and features below, across, and above the bimodality scale.}

\subsubsection{Below the Bimodality Scale}\label{subsub:discuss:belowbimodality}
We consider two scenarios that may explain the valley in $M_{\rm HI,grp}/M_{\rm halo}$ and $\rm FSMGR_{\rm grp}$ for AGN-hosting halos below the bimodality scale: (1) AGN feedback may suppress \HI{} content and star formation, and (2) a reduction in star formation and associated feedback, perhaps for reasons unrelated to AGN fueling, may enable efficient gas inflow to fuel AGN activity or may simply enable AGN detection.

\begin{figure}
    \centering
    \includegraphics{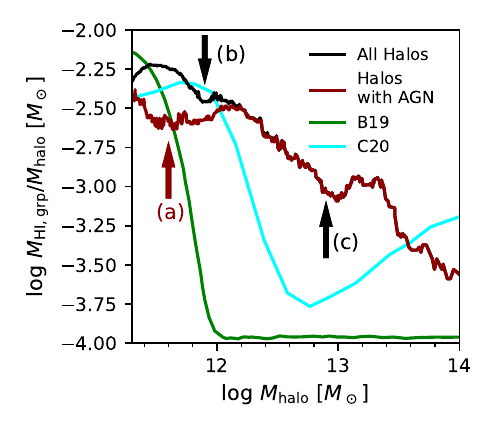}
    \caption{{Observed $M_{\rm HI,grp}/M_{\rm halo}$ versus $M_{\rm halo}$ compared to theoretical predictions. The black line represents the medians  for G3 groups {(excluding groups that lack a definite AGN-hosting or non-AGN-hosting classification)}, calculated in sliding windows as in Figure \ref{fig:agn_hi_sfr}.  The red line shows only AGN-hosting groups. The green and cyan lines show theoretical predictions from \citet{baugh2019galaxy} and \citet{chauhan2020physical}, for which we have normalized their $M_{\rm HI,grp}$-$M_{\rm halo}$ relations to give $M_{\rm HI,grp}/M_{\rm halo}$-$M_{\rm halo}$ relations. Three arrows highlight (a) the broad depression in $M_{\rm HI,grp}/M_{\rm halo}$ for AGN-hosting halos below the bimodality scale, (b) the narrower valley in $M_{\rm HI,grp}/M_{\rm halo}$ for all halos, and (c) a possible high-mass valley, as discussed in \S\ref{subsub:discuss:agn}.}}
    \label{fig:hihmdip}
\end{figure}

In scenario (1), AGN feedback may heat the halo gas or cause atomic gas blowout in galaxies, reducing the fuel supply for star formation. This scenario, gas heating specifically, is predicted by multiple semi-analytic models of the \relation{}, which show a dip at $\log M_{\rm halo} = 10^{12-12.5}\, M_\odot$ \citep{kim2017spatial, baugh2019galaxy, chauhan2020physical}. Prior observational efforts including H23 have failed to detect a dip directly, a failure usually attributed to systematic errors in binning or halo mass estimation \citep{chauhan2021unveiling, dev2023galaxy}. In hindsight, a hint of a dip near $M_{\rm halo} \sim 10^{12}\, M_\odot$ was present in the combined G3 data set in H23 (their Figs. 16 and 18). {Figure \ref{fig:hihmdip} shows how this dip manifests in a plot of $M_{\rm HI,grp}/M_{\rm halo}$ versus $M_{\rm halo}$ along with two theoretical predictions. By comparing the lines for all halos and for AGN-hosting halos, Figure \ref{fig:hihmdip} shows that a shallow valley in the $M_{\rm HI,grp}/M_{\rm halo}$ vs. $M_{\rm halo}$ relation across $M_{\rm halo}\sim 10^{11.7}\, M_\odot$ is caused by a deeper depression in the \HI{} content of low-mass, AGN-hosting halos. Below the bimodality scale, this deeper valley is diluted by the large fraction of non-AGN-hosting halos, so the shallow valley becomes prominent as the fraction of AGN-hosting halos crosses $\sim$50\% around $M_{\rm halo} \sim 10^{11.8}\, M_\odot$ (see Figure \ref{fig:agn_hi_sfr}).} {Figure \ref{fig:hihmdip} also shows that the \citet{baugh2019galaxy} and \citet{chauhan2020physical} SAMs deviate substantially from this observed valley in $M_{\rm HI,grp}/M_{\rm halo}$, predicting much deeper and wider ``troughs'' at higher masses. The observed $M_{\rm halo}\sim 10^{11.8}\, M_\odot$ valley for all halos is much smaller, only $\sim$0.1 dex deep, albeit it is much deeper for AGN-hosting halos taken alone, $\sim$0.25 dex. Also, the observed valley is located mostly \emph{below} the bimodality scale, for both AGN-hosting halos specifically and all halos (Figure \ref{fig:hihmdip}). (We do also see structure in the relation near $M_{\rm halo}\sim 10^{13}\, M_\odot$, which could be a weaker valley similar to that of C20, as will be discussed in \S\ref{subsub:discus:abovebimodality}.) The failure of theoretical models to predict a valley \emph{below} the bimodality scale suggests these models are not yet realistically} including low-metallicity and/or highly star-forming AGN populations such as composite galaxies and {SF-AGN (AGN registering as AGN in the \citealp{veilleux1987spectral} diagnostic plots but as SF in the \citealp{baldwin1981classification} BPT plot; \citealp{polimera2022resolve})}. Composites and SF-AGN dominate the RESOLVE/ECO AGN inventory up to the bimodality scale (\citealp{polimera2022resolve}; M.S. Polimera et al. 2025, in prep.).

In scenario (2), the suppressed \HI{} content of AGN-hosting groups could reflect an inverted causality, wherein lower \HI{} content and thus reduced star formation feedback might allow gas to flow all the way to the black hole, as predicted especially for dwarf galaxies in some models \citep{angles2017gravitational, habouzit2017blossoms, trebitsch2018escape}. In observations, \citet{latimer2019x} have found that AGN are disproportionately uncommon in highly starbursting compact blue dwarfs (see also \citealp{bradford2018effect, penny2018sdss}). Another version of scenario (2) is that reduced SF makes AGN easier to detect \citep{polimera2022resolve}. {The black hole masses of dwarf AGN galaxies are expected to be $\sim 10^{3-5}\,M_\odot$, so their optical AGN signatures can be diluted in the presence of the intense star formation typical of $z\sim 0$ dwarf galaxies \citep{reines2020new}. This bias is reduced in the metallicity-insensitive \citet{veilleux1987spectral} diagnostic plots, which can identify AGN down to 8--16\% AGN spectral contribution for a typical metal-poor dwarf \citep{polimera2022resolve}.}

\subsubsection{Above the Bimodality Scale}\label{subsub:discus:abovebimodality}
{Above the bimodality scale, we see that the trend of higher median $M_{\rm HI,grp}/M_{\rm halo}$ and \fsmgrgrp{} continues to $M_{\rm halo}\sim 10^{12.3}\, M_\odot$, above which non-AGN-hosting halos become nonexistent in our sample. Interestingly, we also find a possible high-mass valley for AGN-hosting halos near $M_{\rm halo}\sim 10^{13}\, M_\odot$ (see Figure \ref{fig:hihmdip}). We leave open the interpretation of this apparent feature, but we note that by analogy with the lower-mass valley, future work may benefit from examining whether the higher-mass feature is associated with any special sub-populations of AGN and their feedback properties. For example, a high-mass valley could hypothetically derive} from a more prominent valley associated with halos hosting AGN in an efficient feedback mode that is diluted by other halos hosting AGN with less intense feedback. The AGN classifications used for our analysis are based on optical emission line diagnostics and mid-IR photometry (\citealp{polimera2022resolve}; M.S. Polimera et al. 2025, in prep.), so we do not have the ability to test such a scenario. Since we lack systematic X-ray and radio AGN classifications, our AGN data will not fully reflect feedback from obscured or radio-loud AGN. Past work suggests that AGN feedback may operate in two modes of growth, ``bright'' and ``radio'' \citep{somerville2008semi,heckman2014coevolution}, both of which could influence halo \HI{} content. In the bright mode, AGN may produce harsh radiation and winds that suppress accretion and star formation. In the radio mode, relativistic jets may heat accreting gas and thereby counteract gas cooling. The numerical abundance of these AGN subtypes may vary with halo mass or galaxy stellar mass. For example, \citet{miraghaei2020effect} showed that the fractions of radio and optical AGN vary with both galaxy stellar mass and environment, considering both field versus group galaxies and centrals versus satellites.  Thus, a careful assessment of AGN subtypes and their corresponding feedback modes would be needed to ascertain how our results relate to AGN feedback.}

\subsubsection{Across the Bimodality Scale}\label{subsub:discuss:acrossbimodality}
As mentioned, {\citet{ellision2019atomic} have observed a reversal across the bimodality scale in whether AGN-hosting or non-AGN-hosting galaxies have higher \HI{} gas-to-stellar mass ratios. {In particular, they found elevated \HI{} content in massive AGN-hosting galaxies ($10 < \log M_*\,\left[M_\odot\right] < 10.8$) at fixed stellar mass. Our results do not show an analogous reversal, but we also lack any non-AGN-hosting halos above $M_{\rm halo}\sim 10^{12.3}\, M_\odot$, illustrating the difference between analyzing individual galaxies and analyzing \emph{entire halos}. With halo mass increasing above the bimodality scale, galaxies increasingly reside in higher richness multiple-galaxy groups, so non-AGN-hosting groups can be scarce even if there are still many individual non-AGN-hosting galaxies. In any case, the reversal \citet{ellision2019atomic} observed} disappeared when the AGN and non-AGN samples were fixed in star formation rate in addition to stellar mass. The authors argued that the original difference stemmed from the fact that, at fixed stellar mass, AGN preferentially reside in star-forming or green-valley galaxies, which tend to have higher \HI{} content. In our case, we have not replicated our Figure \ref{fig:agn_hi_sfr} analysis in fixed bins of $M_{\rm halo}$ and \fsmgrgrp{} together, but we note that the similar behavior of the $M_{\rm HI,grp}/M_{\rm halo}$ and \fsmgrgrp{} lines in Figure \ref{fig:agn_hi_sfr} is expected given the tight correlation between galaxy atomic gas-to-stellar mass ratio and FSMGR \citep{kannappan2013connecting}.}

The question, then, is why AGN are more common in gas-rich and star-forming galaxies. The simple explanation that cold gas fuels both star formation and AGN seems incomplete, since below the bimodality scale, AGN-hosting halos instead have lower \HI{} content and \fsmgrgrp{} than non-AGN-hosting halos. The reversal of this result across the bimodality scale could suggest a relationship to halo gas physics. In the theoretical model of \citet{dashyan2018agn}, there exists a critical halo mass $M_{\rm halo} \sim 10^{12}\,M_\odot$ (at $z\lesssim1$) below which AGN can expel \HI{} gas from dwarf galaxies, and observations have indeed found evidence for \HI{} gas suppression in dwarf galaxy AGN hosts (e.g., \citealp{bradford2018effect}). {This ejective AGN feedback mode contrasts with the  feedback mode that is theoretically expected for halos above the bimodality scale, in which the halo gas, having been shock-heated into a dilute medium, becomes more susceptible to radiative feedback from radio jets \citep{dekel2006galaxy}. Understanding the galaxy mass and/or halo mass dependences of different modes of AGN feedback remains a key research question.}
 
Since AGN feedback efficiency (see \citealp{croton2006many}) controls the location and depth of the dip in the predicted \relation{} of \citet{chauhan2020physical}, such that higher AGN feedback efficiency corresponds to dips located at lower halo masses, the features we observe may help place constraints on the frequency and/or intensity of AGN feedback.}  {The possibility that there are two valleys may imply different regimes of feedback mode and efficiency. We defer quantitative analysis of the location and depth of the valleys to future work. Even interpreting the more convincing $M_{\rm halo} \sim 10^{11.8}\, M_\odot$ valley in this context would require evaluation with mock catalogs since systematic group-finding and halo mass estimation errors affect the shape and scatter of the \relation{} \citep{eckert2017baryonic, hutchens2023resolve, chauhan2021unveiling, dev2023galaxy}.} 

\section{Conclusions}\label{sec:concl}
In this work, we have combined archival RASS X-ray observations with the highly complete and volume-limited RESOLVE and ECO surveys. Using these surveys' G3 group catalogs, comprehensive \HI{} gas and SF data, and state-of-the-art census of optical and mid-IR AGN (including the new SF-AGN category introduced by \citealp{polimera2022resolve} that tracks dwarf AGN), we have examined the connection between group cold gas content and group X-ray emission as a function of halo mass, virialization state, star formation, and AGN presence. Our key results can be summarized as follows:
\begin{itemize}
    \item {We find that as halo mass increases, group \HI{} content decreases while total X-ray emission and hot gas increase. Stacking RASS data, we detect hot gas in groups confidently at $M_{\rm halo} = 10^{12.6-14}\, M_\odot$ and unambiguously at $M_{\rm halo}=10^{14-14.5}\, M_\odot$. We find that the X-ray emission from lower-mass halos can most likely be explained by AGN or X-ray binaries, though we note that the estimation of XRB emission is a key area of uncertainty where future improvements can lead to better measurements of the group hot gas inventory (\S\ref{subsec:discuss:gasinventory}; Figs. \ref{fig:hardband_all_stack}, \ref{fig:cr_vs_mhalo_halomassbins}).}
    
    \item We identify a transition to increased spread in group gas content at fixed halo mass below $\log(t_{\rm cross}/13.8\, {\rm Gyr}) = -0.86$, corresponding to $t_{\rm cross} \sim 2\, {\rm Gyr}$. Defining lower-$t_{\rm cross}$ and higher-$t_{\rm cross}$ categories across this transition, we find that lower-$t_{\rm cross}$ groups show reduced median $M_{\rm HI,grp}/M_{\rm halo}$ compared to higher-$t_{\rm cross}$ groups at fixed halo mass. Above $M_{\rm halo} = 10^{13.3}\, M_\odot$, we additionally find increased X-ray emission for lower-$t_{\rm cross}$ groups. In contrast, we find enhanced X-ray emission for \emph{higher}-$t_{\rm cross}$ groups at $M_{\rm halo} = 10^{12.6-13.3}\, M_\odot$. In this small-group regime, mergers and projection effects are expected to have a larger influence on $t_{\rm cross}$ (\S\ref{subsec:tcrossresults}, \S\ref{subsec:discuss:vir}; Figs. \ref{fig:tcrossmotivation}--\ref{fig:tcross_stacking}).
    
    \item Below the bimodality scale ($M_{\rm halo} = 10^{12.1}\, M_\odot$), halos with AGN exhibit a broad valley in \fsmgrgrp{} and \HI{}-to-halo mass ratio compared to non-AGN-hosting halos at fixed halo mass. We have argued that this result may be consistent with either of two physical scenarios: (1) gas heating or blowout from AGN removes \HI{} and suppresses star formation, or (2) reduced SF feedback enables \HI{} inflow for AGN fueling, or reduced SF simply makes AGN easier to detect (\S\ref{subsec:agnresults}, \S\ref{subsub:discuss:agn}; Figs. \ref{fig:agn_hi_sfr}--\ref{fig:agn_tcross_connection}).
    
    \item {The trend of elevated \HI{}-to-halo mass ratio and \fsmgrgrp{} for non-AGN-hosting halos continues until such halos become nonexistent in our sample at $M_{\rm halo}\gtrsim 10^{12.3}\, M_\odot$, just above the bimodality scale. We also detect significant X-ray emission from AGN-hosting halos at $M_{\rm halo} = 10^{11.5-12.1}\, M_\odot$ and $M_{\rm halo}=10^{12.1-12.6}\, M_\odot$.} {These results hold even if we treat all mid-IR AGN host galaxies as non-AGN-hosting galaxies (\S\ref{subsec:agnresults}, \S\ref{subsub:discuss:agn}; Figs. \ref{fig:agn_hi_sfr}--\ref{fig:agn_tcross_connection}).}
    
    \item {We find that AGN-hosting halos below and just above the bimodality scale show reduced \fsmgrgrp{} at fixed halo mass compared to non-AGN-hosting halos. This pattern for \fsmgrgrp{} matches that for \HI{} content, as expected based on the tight relationship between galaxy FSMGR and galaxy \HI{}-to-stellar mass ratio seen in \citet{kannappan2013connecting} (\S\ref{subsec:agnresults}, \S\ref{subsub:discuss:agn}, Fig. \ref{fig:agn_hi_sfr}).}

    \item Updating H23, we now report evidence for a dip in the \relation{} (in hindsight evident in H23 Figures 16 and 18). This dip {coincides with a shallow valley in the median $M_{\rm HI,grp}/M_{\rm halo}$ vs. $M_{\rm halo}$ relation across $M_{\rm halo} \sim 10^{11.8}\, M_\odot$ as the fraction of halos containing AGN crosses 50\%, reflecting a deeper and wider valley in $M_{\rm HI,grp}/M_{\rm halo}$ for AGN-hosting halos below the bimodality scale.} While an AGN feedback-driven dip in the \relation{} has been theoretically predicted \citep{kim2016spatial, baugh2019galaxy, chauhan2020physical, chauhan2021unveiling}, {these predicted dips correspond to much deeper $M_{\rm HI,grp}/M_{\rm halo}$ troughs at masses above the bimodality scale. The valley we have discovered is centered \textit{between the threshold and bimodality scales} for all halos and is driven by an even broader valley that extends well into the dwarf regime for AGN-hosting halos. We do see a possible high-mass valley in the $M_{\rm HI,grp}/M_{\rm halo}$ vs. $M_{\rm halo}$ relation near $M_{\rm halo}\sim 10^{13}\, M_\odot$. {If it is a valley}, it is much shallower than theory predicts, and its relationship to AGN or AGN subtype activity is {unknown} (\S\ref{subsec:agnresults}, \S\ref{subsub:discuss:agn}; Figs. \ref{fig:agn_hi_sfr}--\ref{fig:hihmdip}).}
\end{itemize}

These results illustrate the wide variety of drivers of group \HI{} content and group X-ray emission at fixed halo mass, thereby deepening our understanding of the rich scatter in the \relation{}, which has only recently been quantified. While our work has clearly demonstrated the importance of virialization state, star formation, and AGN presence in shaping the group gas inventory, it has also raised several interesting puzzles. Some unresolved questions include (a) the physical meaning of the $t_{\rm cross}\sim 2\, {\rm Gyr}$ transition in group gas content; (b) the cause of the reversal of enhanced X-ray emission for lower-$t_{\rm cross}$ vs. higher-$t_{\rm cross}$ groups across the shutdown scale; and {(c) the physical drivers of reduced \HI{} content in AGN-hosting halos, especially in relation to the newly discovered dip in the \relation{} below the bimodality scale.} Future work to address these puzzles will provide essential insights into how galaxies evolve in relation to their dark matter halos.

\begin{acknowledgments}
{We are grateful for feedback from the anonymous reviewer that improved the quality and clarity of this work.} We thank Mugdha Polimera, Andrew Mann, Adrienne Erickcek, Andreas Berlind, Ella Castelloe, Hannah Perkins, {and Chris Richardson} for helpful conversations at various stages of this work. We also thank Michael Corcoran and Koji Mukai for support using HEASARC software/data, including PIMMS. ZLH, SJK, and DSC acknowledge support from NSF grants AST-1814486 and AST-2007351. ZLH and DSC are also grateful for support through North Carolina Space Grant Graduate Research Fellowships.  KMH acknowledges financial support from the grant CEX2021-001131-S funded by MCIN/AEI/ 10.13039/501100011033 from the coordination of the participation in SKA-SPAIN funded by the Ministry of Science and Innovation (MCIN); from grant PID2021-123930OB-C21 funded by MCIN/AEI/ 10.13039/501100011033 by ``ERDF A way of making Europe'' and by the ``European Union''. AJB acknowledges support from NSF grants AST-1814421 and AST-2308161 and from the Radcliffe Institute for Advanced Study at Harvard University. This work has made use of the RESOLVE Survey, which is funded under NSF AST CAREER grant 0955368 (PI S. Kannappan). This research has made use of data and/or software provided by the High Energy Astrophysics Science Archive Research Center (HEASARC), which is a service of the Astrophysics Science Division at NASA/GSFC. This work made use of Astropy ({http://www.astropy.org}) a community-developed core Python package and an ecosystem of tools and resources for astronomy.

\end{acknowledgments}

%

\vspace{5mm}

\bibliography{sample631}{}
\bibliographystyle{aasjournal}



\end{document}